\documentclass[12pt,preprint]{aastex}

\newcommand{\HI}{H~{\sc i}} 
\newcommand{\HII}{H~{\sc ii}}
\newcommand{\kms}{${\rm km~s^{-1}}$}
\slugcomment{{\em To appear in the Astrophysical Journal}}
\shortauthors{McCLURE-GRIFFITHS \& DICKEY} 
\shorttitle{Milky Way Kinematics at the Fourth Quadrant Subcentral Point}

\begin{document} 

\title{Milky Way Kinematics: Measurements at the Subcentral Point of
  the Fourth Quadrant}

\author{N.\ M.\ McClure-Griffiths\altaffilmark{1} 
and John M.\ Dickey\altaffilmark{2}} 

\altaffiltext{1}{Australia Telescope National Facility, CSIRO, PO Box
  76, Epping NSW 1710, Australia; naomi.mcclure-griffiths@csiro.au}

\altaffiltext{2}{School of Mathematics and Physics, University of
  Tasmania, Private Bag 21, Hobart TAS 7001, Australia;
  john.dickey@utas.edu.au}

%--------------------------------------------
\begin{abstract}
%-------------------------------------------
  We use atomic hydrogen (\HI) data from the Southern Galactic Plane
  Survey to study the kinematics of the fourth quadrant of the Milky
  Way.  By measuring the terminal velocity as a function of longitude
  throughout the fourth Galactic quadrant we have derived the most
  densely sampled rotation curve available for the Milky Way between
  $3 \leq R \leq 8$ kpc.  We determine a new joint rotation curve fit
  for the first and fourth quadrants, which can be used for kinematic
  distances interior to the Solar circle. From our data we place new
  limits on the peak to peak variation of streaming motions in the
  fourth quadrant to be $\sim 10$ \kms.  We show that the shape of the
  average \HI\ profile beyond the terminal velocity is consistent with
  gas of three velocity dispersions, a cold component with $\Delta v
  =6.3$ \kms, a warmer component with $\Delta v=12.3$ \kms\ and a fast
  component with $\Delta v=25.9$ \kms.  Examining the widths with
  Galactic radius we find that the narrowest two components show
  little variation with radius and their small scale fluctuations
  track each other very well, suggesting that they share the same
  cloud-to-cloud motions.  The width of the widest component is
  constant until $R<4$ kpc, where it increases sharply.
\end{abstract}

%----------------------------------------------------------
\section{Introduction}
\label{sec:intro}
%----------------------------------------------------------
From the first detection of the atomic hydrogen (\HI) spectral line in
1951, the $\lambda$21-cm line has been used as a probe of the
kinematics of the interstellar medium (ISM) in the Milky Way and the
rotation of the Galaxy.  The rotation curve of the Milky Way has been
a long-standing topic of study with important implications for models
of the structure, mass distribution and dynamics of the Galaxy.  In
one of the earliest studies of the rotation curve, \citet*{kwee54}
noticed irregular depressions in the rotation curve of the first
Galactic quadrant, which they attributed to a lack of \HI\ between
spiral arms.  We now understand that \HI\ emission does not give a
good indication of gas density and therefore the dips do not reflect
density effects, but instead the kinematics of the \HI\ line provide a
direct probe of the dynamical effects of spiral structure.  In
particular, the spiral arms produce deviations, both positive and
negative, away from circular rotation that are responsible for the
irregularities observed in the rotation curves of the inner Galaxy
\citep{kerr62,shane66}.  These departures from circular rotation have
been reasonably well explained by both linear and non-linear spiral
density wave theories \citep{lin69,roberts69}.

Interior to the Solar circle, every line of sight passes a single
subcentral point where it is closest to the Galactic Center and where
its projection on the Galactic plane is tangent to a circular orbit
around the Galactic Center.  At the subcentral point the measured
radial velocity of the \HI\ line reaches its most extreme value,
called the terminal velocity.  In the first Galactic quadrant the
terminal velocity is the most positive velocity in \HI\ spectra,
whereas in the fourth Galactic quadrant it is the most negative
velocity.  By measuring the terminal velocity along all lines of sight
between Galactic longitudes $-90\arcdeg$ and $+90\arcdeg$ it is
possible to derive a measurement of the rotation curve of the inner
Galaxy along the ``locus of subcentral points''.

The shape of the \HI\ velocity profile beyond the terminal velocity
can tell us about the random motions of \HI\ in the Galaxy.  Gas at
these velocities is forbidden by Galactic rotation and therefore gives
us information about the distribution of \HI\ cloud velocities at the
subcentral point.  \citet*{kulkarni85} employed this technique in an
analysis to confirm the existence of two populations of \HI\ clouds,
those with a velocity dispersion on the order of 7-9 \kms\ as found by
\citet{burton76} and \citet{liszt83} and a second population with a
much larger velocity dispersion of $\sim 35$ \kms. The
\citet{kulkarni85} observations towards $l=30\arcdeg$ of a population
of ``fast'' clouds confirmed earlier work by \citet{rad80} and
\citet{anantharamaiah84} that had originally identified this population of
intermediate velocity \HI\ clouds towards the Galactic Center and
toward \HII\ regions, respectively.

This is the first paper in a series about the kinematics of the inner
Milky Way.  In this paper we develop a new technique for measuring the
\HI\ terminal velocity. We apply this technique to new, high
resolution \HI\ data from the Southern Galactic Plane Survey
\citep[SGPS;][]{mcgriff05} to determine the \HI\ terminal velocity
along the locus of subcentral points in the fourth quadrant of the
Milky Way.  The SGPS data have an angular resolution of $130\arcsec$,
which far exceeds the resolution of any previous surveys used to
explore \HI\ kinematics at the terminal velocity.  From the run of
terminal velocities we determine a rotation curve for the inner
Galaxy.  We also use these data to explore the kinematics of the
neutral hydrogen near the subcentral point.  The organization of this
paper is as follows: in \S \ref{sec:termvel} we describe the technique
we have used with the SGPS to determine the terminal velocity curve
and discuss the advantages of using a high resolution survey for
terminal velocity analysis. In \S \ref{subsec:rotcurve} we use the
SGPS terminal velocity curve to derive the rotation curve for the
fourth quadrant and we compare these data with a published rotation
curve of the first quadrant.  In \S \ref{subsubsec:fit} we derive a
new fit to the rotation curve of the inner Galaxy incorporating data
from both the first and fourth Galactic quadrants and in \S
\ref{susubsec:resid} we discuss the magnitude of the velocity
residuals and their relationship to spiral structure.  Finally, in \S
\ref{sec:random} we discuss the shape of the \HI\ velocity profile
beyond the terminal velocity and touch on its implications for the
kinematics of diffuse \HI.

%----------------------------------------------------------
\section{Data}
\label{sec:obs}
%----------------------------------------------------------
The \HI\ data discussed here were obtained as part of the Southern
Galactic Plane Survey \citep[SGPS;][]{mcgriff05}.  The SGPS is an \HI\
spectral line and 21-cm continuum survey of the Galactic plane between
longitudes $253\arcdeg \leq l \leq 358\arcdeg$ (SGPS I) and $5\arcdeg
\leq l \leq 20\arcdeg$ (SGPS II) and with latitude coverage of $|b|
\leq 1\fdg4$.  The observations and analysis of these data are
described in full detail in \citet{mcgriff05}.  The SGPS data were
obtained with the Australia Telescope Compact Array (ATCA) and the
Parkes radio telescope and were combined to provide images with
sensitivity to all angular scales from $130\arcsec$ to several
degrees.  The SGPS I \HI\ data have been released as a series of
overlapping sub-cubes, each covering $11\arcdeg$ of longitude and
$2\fdg89$ of latitude with an angular resolution of 130\arcsec\ and a
spectral resolution of $0.82$ \kms.  The noise in the line free
channels of the \HI\ cubes is 1.6 K.
%----------------------------------------------------------
\section{Measuring the Terminal Velocity}
\label{sec:termvel}
%----------------------------------------------------------
On each line of sight near the Galactic plane circular rotation gives
a velocity relative to the Local Standard of Rest (LSR) that projects
on the line of sight as
\begin{equation}
v_{LSR} = R_{0} \sin{l} \left( \omega - \omega_{0} \right)
\cos{b},
\end{equation}
where $\omega$ is the angular velocity of Galactic rotation at the
point of interest, and $\omega_{0}$ is the corresponding Galactic
rotation of the solar circle ($R = R_{0}$). Throughout this paper we
adopt the IAU standard values for the distance to the Galactic Center,
$R_0=8.5$ kpc, and $\omega_0=\Theta_0 / R_0 = 220~{\rm km~s^{-1}}/8.5$
kpc. The terminal velocity is defined as the extreme velocity on any
line of sight at $b\approx0\arcdeg$, which occurs at the subcentral or
tangent point where the galactocentric radius is $R=R_{0} \sin{l}$ for
any realistic rotation curve.  The projected radial component of this
maximum allowed circular rotation velocity gives the rotation curve of
the Galaxy, $\omega(R)$, or alternatively $\Theta(R)$.  At this
position the helio-centric distance, $r$, is determined from geometry
simply as $r=R_0 \cos l$.

For cold gas in a perfectly circular orbit, and in the absence of
turbulent motions, the \HI\ spectrum would drop to zero at velocities
beyond the terminal velocity (in the fourth Galactic quadrant: at more
negative velocities).  However, \HI\ and other real spectral line
tracers do not drop discontinuously to zero for velocities beyond the
terminal velocity because the gas has a measurable thermal velocity
width as well as bulk random motions.  For most of the gas in the
Milky Way these random motions are roughly 5\% or less of the circular
velocity \citep{burton76}, so line profiles near latitude zero drop
quite sharply near the terminal velocity.  Analysis of the shape of
the \HI\ velocity profile beyond the terminal velocity can give
us information about the random motions of the gas in the region near
the subcentral point.  Similarly, understanding the distribution of
random velocities helps refine the method of estimating the extreme
allowed circular rotation velocity from the line profiles.  Here we
carefully fit the terminal velocity profile to reveal both.

A variety of methods have been employed to define a point along the
velocity profile as the terminal velocity.  For example,
\citet{kerr62} defined the terminal velocity, $V_{LSR}$, as the last
peak of the profile, whereas \citet{sinha78} used a relative
definition in which the terminal velocity is where the line profile
crosses a threshold set at half the value of the most extreme allowed
velocity emission peak.  \citet*{shane66} and \citet*{celnik79}
derived functional forms for the shape of the \HI\ profile and the
terminal velocity and fit their model profiles with a defined terminal
velocity to the observed profiles to define the exact drop off.  More
recently, \citet{malhotra95} used an absolute brightness temperature
threshold to define the terminal velocity.  We have employed a
compromise technique, which starts with an absolute brightness
temperature threshold to adjust the velocity profile followed with
function fitting to define the exact velocity of the underlying
discontinuity, independent of threshold choice.  We made two passes
through the data in an effort to avoid the effects of an arbitrary
choice of measurement parameters.

\subsection{Continuum Blanking}
An important advantage of a high angular resolution survey like the
SGPS is that the effects of absorption of continuum emission from
beyond the subcentral point can be removed from the data.  When
observed toward a bright continuum source, the \HI\ spectrum shows a
mixture of emission and absorption at all velocities.  The absorption
can overwhelm the emission, particularly near the subcentral point
where velocity crowding causes a long segment of the line of sight
to pile up in the spectrum near the terminal velocity.  For a low
resolution survey, such as one observed with a single dish telescope,
absorbed and non-absorbed pixels are averaged with a weighting given
by the beam shape.

\begin{figure}
\centering
\includegraphics[width={\textwidth}]{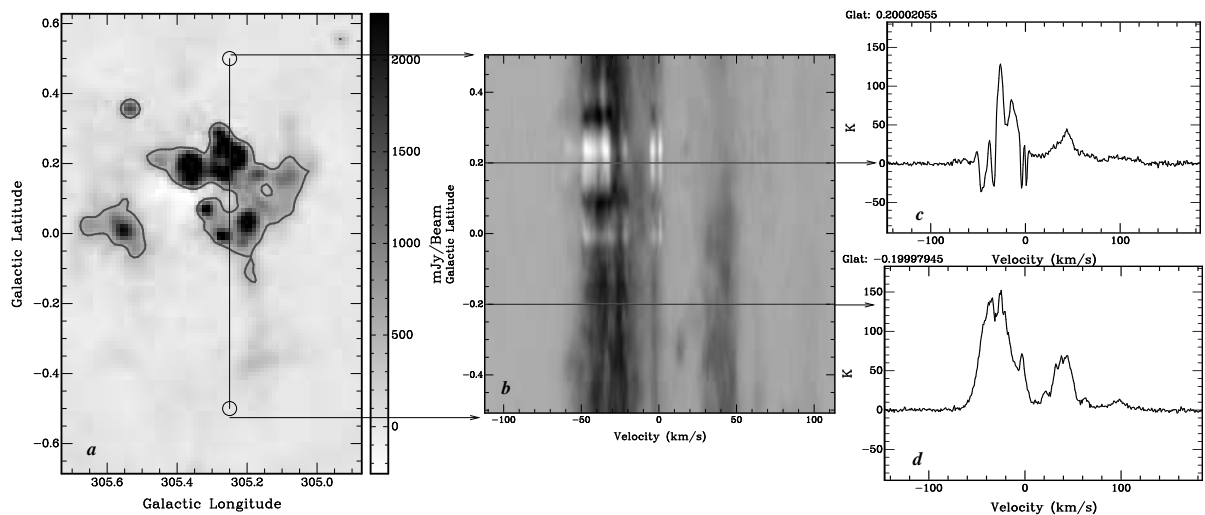}
\caption[]{An illustration of the effect of continuum absorption on
  the \HI\ spectra. The left hand panel shows a grey scale map of
  continuum sources near longitude $l=305\fdg25$. The vertical line
  spanning $+0\fdg5 \geq b \geq -0\fdg5$ shows the area covered in a
  single latitude average.  The latitude-velocity plot in the center
  panel shows the \HI\ line intensity along this line, with velocity
  as the horizontal axis and grey scale indicating brightness
  temperature.  The white areas have $T_B < 0$ because the continuum
  has been subtracted from the spectral line cubes.  The right hand
  panels show two spectra, at $b=+0\fdg2$ and $b=-0\fdg2$, sliced from
  the latitude velocity plot.  The upper one is polluted by absorption
  toward the continuum and must be excluded from the calculations.
\label{fig:jd1}}
\end{figure}

The blended emission and absorption in a low resolution survey leads
to an underestimate of the maximum velocity, as absorption towards
continuum sources in the beam will tend to flatten the steep ``cliff
edge'' of the emission profile near the terminal velocity.  A clear
example of this effect is shown in Figure~\ref{fig:jd1}, which shows
two spectra at the same longitude at two different latitudes (b=$\pm
0\fdg2, l=305\fdg25$) where the first has bright continuum in the
background and the second does not.  The effect of the absorption is
to reduce the height of the emission, in extreme cases the brightness
temperature can even go negative because the continuum has been
subtracted from the SGPS data.  This causes the apparent rotation
velocity, whether defined by a threshold or by profile fitting, to be
biased to unrealistically low values wherever there is bright
background continuum emission.  To avoid this bias in our analysis we
have used the SGPS continuum images \citep{haverkorn06} to blank all
pixels with $T_{cont} > 0.5$ K.  Overall we blank about 10\% of the
survey area.  Without blanking we find that the measured terminal
velocities show more scatter from one longitude to the next.  Single
dish surveys are not able to do this because their much larger beam
size irrevocably blends the spectra toward continuum sources with the
surrounding emission spectra, reducing the measured terminal velocity.

\subsection{Thresholding on Latitude Averages}
After continuum blanking our first step is to chose a brightness
temperature threshold that selects a velocity that is part way down
the steepest slope in the emission profile, the ``cliff'' just past
the subcentral point velocity.  An individual spectrum is rather noisy
($\sigma _{T}$ = 1.6 K) so it is necessary to perform some kind of
latitude average before thresholding.  A simple average over latitude
would return the mean of this distribution.  Alternatively we could
work with the median, i.e.\ the 50th percentile $T_b$, or some other
value of $T_b$ defined from this distribution such as the 25th or 75th
percentile.  Because all latitudes are included in constructing the
spectrum any of these statistics could be used on the distribution
of $T_b$ at each velocity to construct a spectrum that has relatively
low noise.

The SGPS interferometer data latitude coverage of $\pm 1\arcdeg$
corresponds to 55 times the beam width, or about 70 independent points
for each beam width in longitude.  The scale height, $h$, of the \HI\
gas in the inner Galaxy is about $\pm 150$ pc in $z$ \citep{dickey90},
which gives $\pm 1\arcdeg$ angular width at the nominal galactocentric
distance $R_{0}=8.5$ kpc.  Thus all lines of sight with $|b| < 0\fdg5$
pass the subcentral point when they are still well inside the \HI\
layer ($|z|\ll h$), so all spectra at the same longitude show roughly
the same sharp drop beyond the subcentral point velocity.  Corrections
to the projected rotation velocity to account for the non-zero
latitude, both due to the reduced projection of the circular rotation
velocity vector along the line of sight and due to departures from
cylindrical rotation of the gas above or below the mid-plane at the
subcentral point, are insignificant ($\sim 10^{-3}$) for latitudes
$|b|<1\arcdeg$.  For the terminal velocity analysis
described here we have used only spectra within $|b| <0\fdg5$, so the
effect of the finite scale height is very slight.  We can therefore
consider any set of spectra at the same longitude as realizations from
the same statistical ensemble, differing in their profile shapes due
to small scale random motions near the subcentral point, but all
having the same underlying rotation velocity.

\begin{figure}
\centering
\includegraphics[angle=-90,width={\textwidth}]{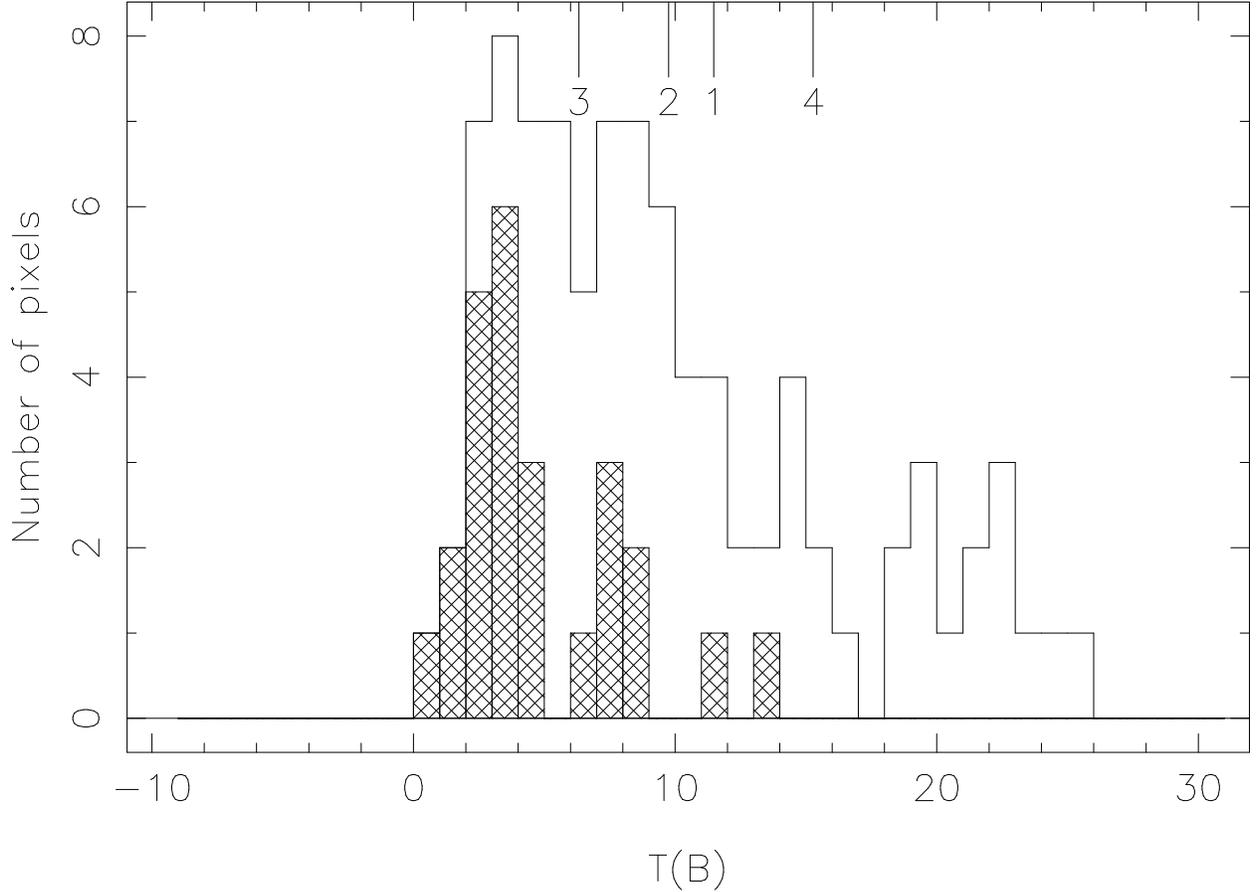}
\caption[]{Distribution function of the brightness temperature
  measured at $v=-61.83$ \kms, $l=305\fdg25$, for
  latitudes $+0\fdg5 \geq b \geq -0\fdg5$, which is the range shown on
  Figure~\ref{fig:jd1}.  The velocity displayed here is the threshold
  crossing velocity for this longitude, meaning it is the most
  negative velocity for which
  the mean of the brightness temperature distribution is above the 10 K threshold.  The mean,
  median, 25th percentile, and 75th percentile of the distribution are
  indicated by the marks labelled 1, 2, 3, and 4 respectively. The
  shaded pixels are eliminated from the distribution before computing
  the statistics because they correspond to positions where the
  continuum is above the cutoff of 0.5 K.  Distribution functions like
  this are analysed at every longitude, and the four statistics are
  used to compute four alternative velocity shifts, $v_t(l)$, where
  each one crosses the 10 K threshold.
  \label{fig:jd2}}
\end{figure}

For each longitude we have used the many independent spectra at
different latitudes to study the distribution function of \HI\
emission brightness temperature, $T_b$, at each velocity.  An example
of one such distribution function of $T_b$ at $-0\fdg5 < b < +0\fdg5$,
$l=305\fdg25$ and $v=-61.83$ \kms\ is shown in Figure~\ref{fig:jd2}.
This is the velocity where the mean of the distribution crosses the 10
K threshold, i.e.\ the most negative velocity with mean $T_b(v)>$ 10
K.  The positions of all four statistics, mean, median, 25th and 75th
percentile are marked on Figure \ref{fig:jd2}.  We have used all four
statistics to determine the velocity of threshold crossing.  We have
carried the velocities found with each statistic through the second
part of the analysis, as discussed in \S \ref{subsec:errfunc}, to
establish which statistic is most robust.

\begin{figure}
\centering
\includegraphics[height=7in]{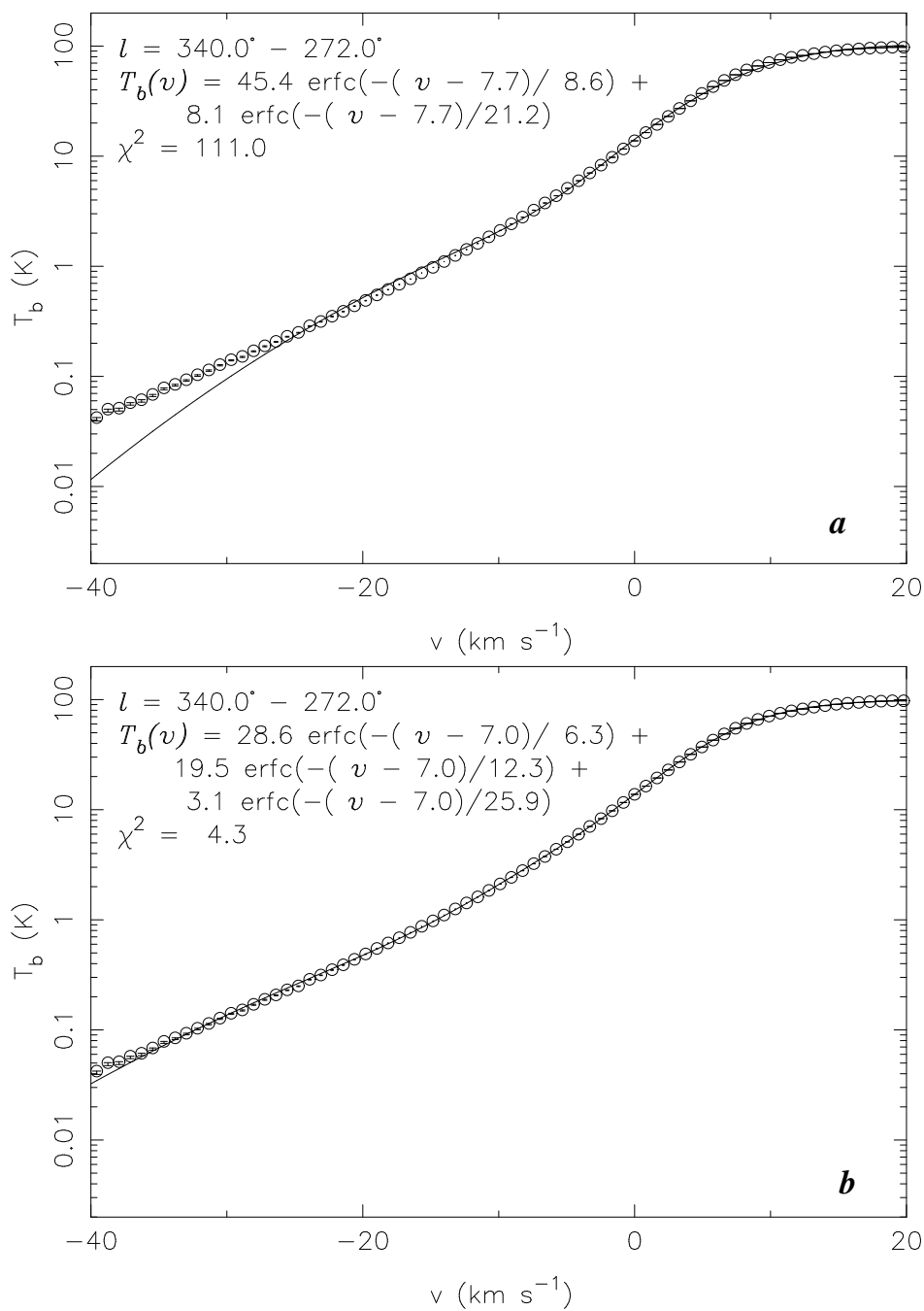}
\caption[]{Error function fits to the longitude averaged terminal
  velocity profile.  Panel {\em a} shows a two component fit to the
  profile while panel {\em b} shows a three component fit to the
  profile.  Clearly the three component fit is much better able to
  account for the tail of the profile.
\label{fig:errfav}}
\end{figure}

Whatever the choice of threshold in $T_b$ and the statistic chosen to
determine the average $T_b(l,v)$ from $T_b(l,b,v)$, after continuum
blanking and averaging we have a function $v_t (l)$, which is the
velocity of threshold crossing.  This is not the rotation curve, but
it is the first step toward the rotation curve.  Using this function,
we shifted all spectra in the survey by the velocity corresponding to
their longitude, $v_t(l)$, and then computed an {\em l-v} diagram
based on the average in latitude.  By shifting the spectra in velocity
by $v_t(l)$ we removed the gross effects of Galactic rotation so that we
could average spectra in longitude. Figure \ref{fig:errfav} is a
terminal velocity profile averaged over the longitude range
$272\arcdeg \leq l \leq 340\arcdeg$.  The velocity scale of Figure
\ref{fig:errfav} reflects the shift so that 0 \kms\ corresponds to the
threshold crossing, $v_t(l)$.  The subcentral point is at a positive
velocity somewhat offset from $v_t$.  The value of the offset depends
on the threshold choice, and the choice of statistic (mean, median, or
other percentile) used for the latitude averaging.  In the next
section we describe how we determine the offset to the subcentral
point by fitting complementary error functions, ${\rm erfc}(v)$, to
the terminal velocity profile.

%----------------------------------------------------------
\subsection{Error Function Fitting of the Terminal Velocity Shape}
\label{subsec:errfunc}
%----------------------------------------------------------
The second step in determining the terminal velocity was to fit the
shifted spectral averages with a functional form to establish a robust
velocity for the subcentral point.  For cold gas with no velocity
dispersion the \HI\ profile would drop sharply at the terminal
velocity.  However, the \HI\ profile does not show the ideal edge of a
Heaviside step function, but is instead smoothed out by the velocity
dispersion of the gas.  A commonly used technique to find the edge of
a step function blended by some natural dispersion is to fit a
complementary error function, ${\rm erfc}$, which returns the position
of the edge and as well as the width of the blending process.  The
relationship between the step function and the error function can be
seen clearly if we consider the Heaviside step function as the limit
of a complementary error function
\begin{equation}
H_{v_o}(v) = \frac{1}{2}\lim_{\Delta v\to0}\ {\rm erfc}((v_o-v)/\Delta v),
\end{equation}
where $v_o$ is the velocity offset of the edge and $\Delta v$
characterizes the deviation from an ideal edge.  The use of
an error function to fit the \HI\ profile beyond the terminal velocity
is not without precedent, \citet{kulkarni85} used this form in their
exploration of the velocity dispersion of \HI.  In this application,
the width $\Delta v$, relates to the Gaussian dispersion as $\Delta v
=\sqrt{2} \sigma_v$.  

There is also some physical justification for using an error function
to fit the \HI\ line profile.  Beyond the terminal velocity,
$v_{LSR}$, we expect that the gas must become optically thin, meaning
that the brightness temperature becomes proportional to the column
density of atoms in each velocity interval, $\delta v$.  If
the density of gas at allowed velocities is a constant, $n_o$, then
the brightness temperature at a velocity beyond the terminal velocity
($v < v_o$, where $v_0<0$ for the fourth quadrant) for which the gas
is optically thin is
\begin{equation} T_b(v) \ \delta v \ = \  \frac{n_o}{C_o} \
  \int_{-v}^{\infty} \left| \frac{dv^{\prime}}{dr} \right|^{-1}
  f(v-v^{\prime}) dv^{\prime}, 
\label{eq:tb}
\end{equation}
where the upper limit on the velocity integral is not really infinity,
but the range of allowed velocities is so wide on most lines of sight
that the upper limit does not matter.  The function,
$f(v-v^{\prime})$, is the velocity distribution function of \HI\ atoms
contributing to the emission, assumed to be a Gaussian.  The constant
$C_o$ is the familiar $1.823\times 10^{18}~{\rm cm^{-2}}$ (K km
s$^{-1}$)$^{-1}$ conversion factor between brightness temperature and
column density for the $\lambda$21-cm line.  Finally, $\left|dv^{\prime}/dr
\right|$ is the magnitude of the inverse of the velocity gradient
along the line of sight, evaluated at position $r$ that has radial
velocity $v^{\prime}$. As we can see, Equation~\ref{eq:tb} bears a
strong resemblance to the complementary error function but for the
$\left| dv/dr\right|$ term inside the integral.  \citet{celnik79}
showed that for a smooth rotation curve the velocity gradient,
$dv/dr$, near the terminal velocity gradient is small and directly
proportional to distance, $r$ with a slope given by the local value of
Oort's constant over $R$.  At the subcentral point the velocity
gradient crosses zero, such that the absolute value of the velocity
gradient has a sharp discontinuity at the subcentral point.  For
optical depths $\tau \ll 1$, this sharp discontinuity should result in
a sharp peak in the \HI\ profile at the terminal velocity.  In fact,
we generally do not observe a sharp peak, but a smoothly rolled off
velocity profile, which suggests saturation in the line profile near
the subcentral point.  The effects of saturation and spatial variations
in the density appear to mask any correlation between the magnitude of
the velocity gradient and the amplitude of the emission at the
terminal velocity, such that the erfc function scales to connect
smoothly with $T_b(v_o)$, whatever it is.  Thus if we make the
assumption that for low longitudes the term $n_o \left|dv^{\prime}/dv
\right|^{-1}$ is roughly a constant near the subcentral point then it can be taken
out of the integral, leaving the complementary error function for the
velocity dependence of the line profile beyond the terminal velocity.
Fitting this function reveals both the terminal velocity and the width
of the underlying Gaussian distribution of velocities.

As was pointed out by \citet{kulkarni85}, a single Gaussian for $f(v)$
is not a good fit to the data.  A single Gaussian is unable
to to fit the sharp drop plus a long tail of emission beyond the
terminal velocity.  The next most simple functional
description of $T_b(v)$ is the sum of two or three complementary
error functions, corresponding to two or three Gaussians of widths,
$\sigma_v$:
\begin{equation}
  T_b(v) = \sum_{i=1}^3 \, a_i\, {\rm erfc}(-1*(v-v_o)/\Delta v_i),
\label{eq:errfunc}
\end{equation}  
where the width of the error functions, $\Delta v_i$, is related to
the original Gaussian width as $\Delta v_i = \sqrt{2}\sigma_{v_i}$.  The
top and bottom panels of Figure~\ref{fig:errfav} show fits of two and
three error functions, respectively, to the longitude average of the
shifted $T_b(v)$.  Clearly the low-level emission tail cannot be well
fitted by even the sum of two error functions, instead it is necessary
to use a sum of three error functions.  We discuss the implications of
this for \HI\ kinematics in \S\ref{sec:random} below.

To determine the true terminal velocity at the subcentral point
$v_{LSR} = v_{t} + v_o$, we fit error functions to shifted and
longitude averaged $T_b(v)$ profiles to return values for
$v_o(l)$. The formal fits were done by minimizing the chi-squared for
the fit using Powell's method of minimization \citep{press92}. We
found that $v_o(l)$ determined from fitting a sum of two error
functions was identical to within 2\% to $v_o(l)$ determined from
fitting a sum of three error functions.  Therefore, for speed and
robustness against noisy spectra we determined $v_o(l)$ by fitting the
sum of only two error functions.

The fitted values for $v_o(l)$ depend on the threshold value and the
statistic used to average the spectra in latitude, but the true
terminal velocity, $v_{LSR}$, is largely independent of threshold and
statistic. This is illustrated in Figures~\ref{fig:plotrot} and
\ref{fig:plotrot2}.  Figure~\ref{fig:plotrot} shows the threshold
crossing velocities, $v_t$, before and after correction by $v_o$ for
twenty longitude bins between $l=340\arcdeg$ and $l=292\arcdeg$.  The
four crosses in each longitude bin are the values of $v_{t}$ obtained
from the 10 K threshold crossing for each of these four statistics.
In each case the outliers are the 25th and 75th percentiles, whereas
the median and mean statistics give similar results.  Fitting the
velocity offset, $v_o$, to the corresponding spectrum averages gives
different values that usually make up for the difference in the
percentiles, so that the sum, $v_{LSR}=v_{t} + v_o$ comes out the same
for all four statistics.  The dotted curve corresponds to $v_{LSR}$
for the 75th percentile, the dashed curve for the 25th percentile, and
the two solid curves for the mean and median.  For longitudes $l \leq
325\arcdeg$ the scatter among the four values of $v_{t}$ is 1 to 2
\kms, but the scatter among the four values of $v_{LSR}$ is only 0.26
\kms.  In this range, the continuous curves on
Figure~\ref{fig:plotrot} are indistinguishable.  For longitudes $
327\fdg5 \leq l \leq 330\arcdeg$ the points spread apart, but the
values of $v_{LSR}$ have less scatter than $v_{t}$.  At all longitudes
the values of $v_{LSR}$ obtained from the mean and median in latitude
agree to better than 0.6 \kms, except at $l=328\fdg25$ where their
difference is 4.6 \kms\ and at $l=338\fdg25$ where the difference is
1.2 \kms.  The striking agreement of the values of $v_{LSR}$ obtained
from the different choices of statistics to define $v_{t}$ after
correction for the offset velocity, $v_o$, demonstrates that the erfc
fitting effectively removes any bias introduced by the arbitrary
choice of statistic for the latitude averaging.

The velocity offset correction also effectively corrects for any bias
introduced by the choice of threshold values.  This is shown in
Figure~\ref{fig:plotrot2} where we compare $v_t$ and $v_{LSR}$, both
derived using longitude averages but with brightness thresholds of 10
K and 20 K. The 20 K threshold reduces the velocity shifts by 3-5
\kms, but this is made up by smaller offsets between the threshold
crossing velocity and the edge of the underlying step in the
brightness distribution derived from the error function fitting, so
that the discrepancy in $v_{LSR}$ is much smaller.  The difference in
$v_{LSR}$ derived using the two thresholds is less than 1 \kms\ for
longitudes $l \leq 325\arcdeg$, and less than $2.5$ \kms\ for
longitudes $l \geq 332\fdg5$. Overall the adjustment from the erfc
fitting does a good job of correcting for bias due to the arbitrary
choice of threshold.  The difference in threshold crossing velocity,
$v_t$, is typically 5 to 15 \kms\ before applying the shift, $v_o$,
obtained from the erfc fits.

As seen in Eq.\ \ref{eq:tb} and discussed above, the brightness
temperature is related to product of the local \HI\ density, $n_o$ and
the inverse of the velocity gradient, $|dv/dr|^{-1}$.  Because of this
relationship there is a danger with a pure threshold method that a
localised change in $n_o$ or $dv/dr$ could increase the local
brightness temperature and hence alter the measured terminal velocity.  The
velocity offset correction method that we employ provides some
robustness against masking density or velocity gradient changes as
terminal velocity changes by compensating for variations in the
threshold velocity. Our estimates of the offset velocity made by
fitting error functions to the terminal velocity profile are dominated
by the velocity, rather than the brightness temperature, of the peak
of the error function at $v\approx v_{LSR}$.  Therefore, the value of
the threshold, or even its position along the velocity tail, is
largely irrelevant to our determination of $v_{LSR}$ and we are immune
to density and velocity gradient effects on $T_B$.  It is worth noting
however, that the offset fitting method does not provide much
protection against very large-scale (factor of $\sim 2$ or more)
variations in $n_o$ or $dv/dr$ away from the subcentral point which
can effectively shift the velocity of the peak back from the terminal
velocity.  At present there is no method that can resolve these from
true variations in the rotational velocity and it remains a limitation
of all Milky Way rotation curves.

Thus the final values of $v_{LSR}$ should be repeatable and
independent of the choice of threshold or statistic to within 0.5
\kms\ for $l \leq 325\arcdeg$, and within 3 \kms\ for $l \geq
332\fdg5$.  In the range $326\fdg5 \leq l \leq 332\fdg5$ there is a
discrepancy in the results from the two thresholds that is as large as
10 \kms. Similarly, the range $327\fdg5 \leq l \leq 330\arcdeg$ gives
a large variation in threshold crossing velocity using the same
threshold but different statistics in latitude (see
Figure~\ref{fig:plotrot}).  For example, the mean and median for the
10 K threshold give $v_t=-143.1$ \kms\ and $v_t=-138.3$ \kms,
respectively.  The offsets derived from the error function fitting in
that case are $v_o= 24.5$ \kms\ and $v_o=23.7$ \kms, which reduce the
discrepancy, but only slightly.  This appears to be due to abnormal
variation in the slope of the profiles beyond the terminal velocity in
this region, making the method of latitude averaging, and the choice
of threshold, give different results.  This is possibly evidence for
heightened rate of stirring up random motions in the medium, due to
increased star formation activity in the galactocentric radius range
$3.9 \leq R \leq 4.5$ kpc.  As discussed below, this is a region where
the rotation velocity increases rapidly with R, which may be due to
non-circular motions associated with a spiral arm.

%----------------------------------------------------------
\section{The Terminal Velocity Curve}
\label{sec:rotation}
\begin{figure}
\centering
\includegraphics[width={\textwidth}]{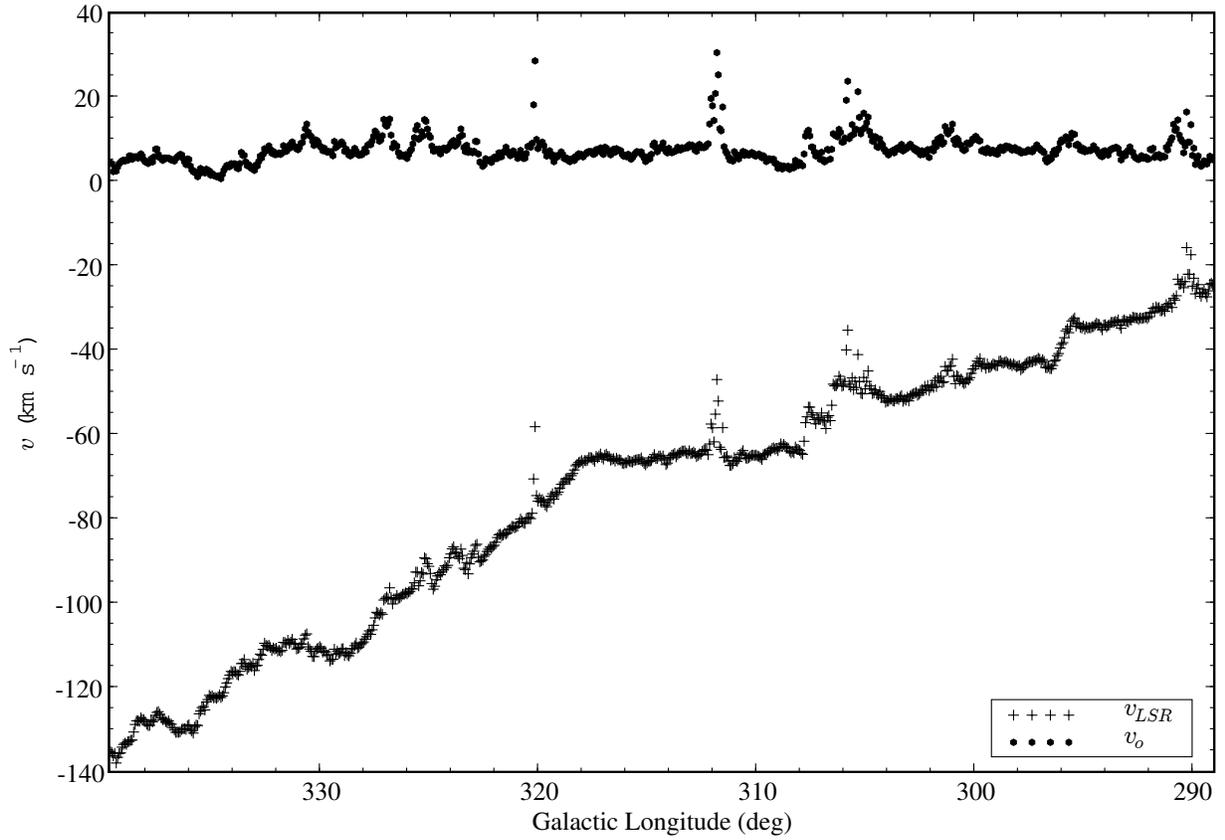}
\caption[]{\HI\ terminal velocities, $v_{LSR}$, and the offset velocity,
  $v_o$, versus Galactic longitude for the fourth
  quadrant of the Milky Way.  
  \label{fig:termcurve}}
\end{figure}

%----------------------------------------------------------
Using the threshold method and error function fitting we have
determined the terminal velocity, $v_{LSR}$, for 1000 longitude bins
of width $3\arcmin$ between $l=339\fdg5$ and $l=272\arcdeg$.  We used
a threshold value of 10 K to determine $v_t(l)$ and we fit the sum of
two error functions for each longitude bin to determine, $v_o(l)$.
Table \ref{tab:termvels} gives the measured terminal velocity,
$v_{LSR}$, and central longitude, $l$, for each longitude bin.  The
terminal velocity curve, together with the values for $v_o(l)$, is
shown in Figure~\ref{fig:termcurve}.  For clarity we have not plotted
the error bars, which should reflect the estimation of the error from
on thresholding and fitting described in \S \ref{subsec:errfunc}.

The longitude range used for this terminal velocity curve analysis has
been chosen to avoid extreme non-circular motions associated with the
inner 3 kpc of the Galaxy.  In order to translate the measured values
of $v_{LSR}$ to a rotation curve we must assume circular rotation
along the locus of subcentral points.  We know, however, that circular
rotation breaks down in most of the very inner Galaxy. At $R\lesssim3$
kpc, gas orbits become elliptical because of the dynamical effects of
the three kiloparsec arm and the bar \citep{burton93,ferriere07}.  The
dynamics of the \HI\ in the innermost region of the Galaxy will be the
subject of a future paper and will not be discussed further here.

Automated function fitting produced spurious fits for $\sim 2$\%
of the longitude bins.  Some of these are visible as cusps in the
terminal velocity curve at $l=306\arcdeg$, $l=312\arcdeg$ and
$l=320\arcdeg$ in Fig.\ \ref{fig:termcurve}. Examining the \HI\
datacubes we found that at these three longitudes there are discrete
\HI\ clouds 10-20 \kms\ beyond the terminal velocity, which disrupt
the fitting.  These clouds are spatially only marginally resolved, have low
brightness temperature features ($T_b\sim 10$ K) and FWHM velocity
widths of $\sim 10$ \kms.  They may be similar to the outlying clouds
described by \citet{stil06}.  In the remainder of the SGPS \HI\
profiles used in this analysis the high velocity linewings are smooth
and do not show small, discrete components.  To remove the velocity
spikes at $l=306\arcdeg,312\arcdeg$ and $320\arcdeg$ from the
subsequent analysis we have run a median filter with a width of 9
samples across the $v_o(l)$ values.

\subsection{Comparison with CO terminal velocities}
\HI\ because of its ubiquity is often hailed as an excellent tracer of
the terminal velocity curve.  However, its ubiquity together with
large ($\sim 5-10$ \kms) thermal and turbulent line widths can lead to
line blending.  Because of these limitations, the terminal velocity
curve has often been studied from observations of the $^{12}$CO
$J=1-0$ transition. Unlike \HI, the thermal linewidth of the $^{12}$CO
line is typically very small, $< 1$ \kms\ and even the turbulent width
is typically only $\sim 2$ \kms\ \citep{burton76}, which reduces the
problems of line blending.  Because of this $^{12}$CO determined
terminal velocity curves should demonstrate less scatter than their
\HI\ counterparts.  Comparison of $^{12}$CO and \HI\ terminal
velocities can test how free our \HI\ terminal velocities are from
systematic effects caused by turbulence and line-blending.  In Figure
\ref{fig:vsinl} we have overlaid values for the recent $^{12}$CO
terminal velocity curve measured by \citet{luna06} on our \HI\
terminal velocities as a function of $\sin{l}$.  The \citet{luna06}
values are sampled every $0\fdg125$ and were selected as the
half-intensity point of the extreme velocity peak.

The agreement between the \HI\ and $^{12}$CO datasets is striking; the
two curves traces each other extremely well including many of the dips
and wiggles.  On average the scatter in the $^{12}$CO data points is
about three times that of the \HI\ data points.  The larger scatter in
the CO data may be partially attributed to the fact that these data
are less sensitive than the \HI\ data and also that \citet{luna06}
used a threshold method for determining the terminal velocity.
However, it is likely that physical effects relating to the small
volume filling factor of CO contribute significantly to the difference
in scatter between \HI\ and CO. The low filling factor of CO means
that observations at the subcentral point typically detect a very
small number of discrete CO clouds.  The CO observations are therefore
sensitive to the individual velocity dispersions of the CO clouds,
rather than an ensemble average as measured with \HI.  This can lead
to a larger scatter for individual measurements.  The comparison of
these two curves raises confidence that \HI\ is not overly dominated
by blending or turbulence and may in fact indicate that \HI\ is a more
robust measure of the terminal velocity.

\subsection{Determining {\boldmath $AR_0$}}
\begin{figure}
\centering
\includegraphics[width={\textwidth}]{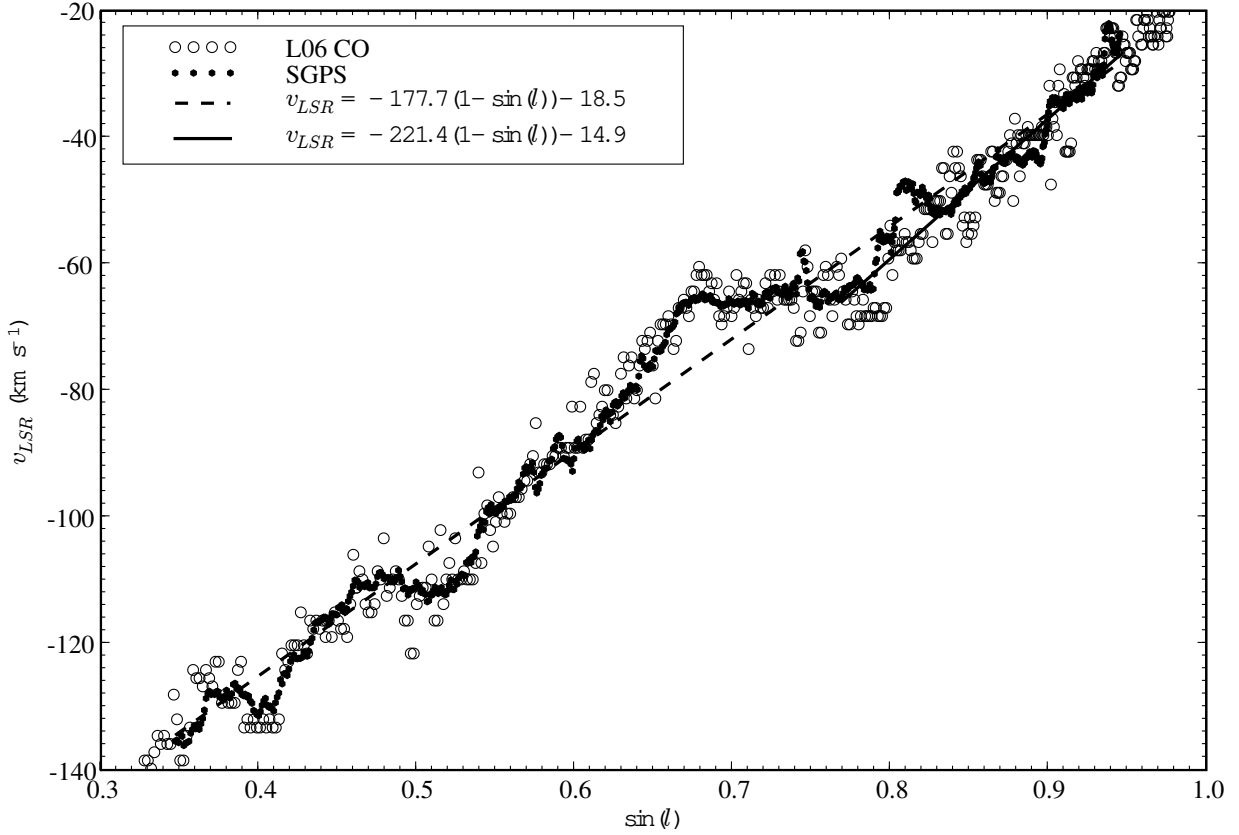}
\caption[]{\HI\ and CO terminal velocities versus $|\sin l|$ for the
  fourth quadrant of the Milky Way.  CO terminal velocities are from
  \citet{luna06}[L06].  The \HI\ terminal velocities have been fit with a linear function of
  the form $v_{LSR} = \Theta_0(| \sin l| -1) +v^{\prime}$.   The slope of the fit is
  dependent on the range of $\sin l$ used in the fitting.  The dashed
  line shows a fit over the range $0.35 \leq |\sin l| \leq 0.95$ and
  the solid line shows the fit over the restricted range $0.8 \leq |\sin l| \leq 0.95$. The
  fitted values are given in the legend. 
  \label{fig:vsinl}}
\end{figure}

Oort's $A$ constant is an important quantity for probing differential
Galactic rotation in the vicinity of the Sun.  Although \HI\ data are
not able to determine $A$ directly, terminal velocities have been
used to measure the product $AR_0$, where
\begin{equation} 
v_{LSR} = -2AR_0(1-\left| \sin{l} \right|) + v^{\prime}.
\label{eq:velfit}
\end{equation}  
measured near to the Sun \citep[e.g][]{gunn79}.  In principle we could
determine $AR_0$ by simply fitting Equation \ref{eq:velfit} to the
SGPS terminal velocities as a function $\sin{l}$ plotted in
Figure~\ref{fig:vsinl}.  This is best done near the Sun because $A$ is
a locally determined quantity.  However, this is complicated
enormously by streaming motions both near the Sun and throughout the
inner Galaxy, which lead us to fit over large ranges of $\sin{l}$ in
order to overcome localized non-circular motions.  \citet*{lin78} use
nonlinear density wave theory to demonstrate that the value obtained
for $AR_0$ depends strongly on the location where it is measured.
From Figure~\ref{fig:vsinl} it is clear that although the general
trend for the SGPS terminal velocity is linear, there are significant
departures from a linear fit. We found that the values determined for
$AR_0$ depend on the range of longitudes used in the fitting.  We
immediately eliminated data in the range $\sin{l}>0.95$ because the
magnitude of the measured terminal velocities at these longitudes
($\sim 10$ \kms) is comparable to the width of the terminal velocity
profile as well as any expected deviations due to streaming motions.
In Figure \ref{fig:vsinl} we show two fits to the data over different
ranges of $\sin{l}$.  We found that fits to $0.8 \leq \sin{l} \leq
0.95$ return $AR_0=110\pm 2$ \kms, which agrees very well with
\citet{gunn79} and implies $A=13~{\rm km~s^{-1}~kpc^{-1}}$ if
$R_0=8.5$ kpc.  However, fits over a broader range of $\sin{l}$
extending up to $\sin{l}=0.35$ return $AR_0=89\pm 4~{\rm km~s^{-1}}$,
which implies $A=10.5~{\rm km~s^{-1}~kpc^{-1}}$ if $R_0=8.5$ kpc.
\citet*{alvarez90} found a similar difference between the slope for
fits to data up to $l=320\arcdeg$ and for data up to $l=340\arcdeg$
and attributed the difference to anomalous velocities in Carina, as
discussed by \citet{humphreys74}.  The large range of $AR_0$
determined within the same data-set reveals just how sensitive the
product of these constants is to the location in which it is measured.
As warned by \citet{lin78} one should exercise extreme caution when
determining $AR_0$ from \HI\ or CO terminal velocity measurements.

Recognising the intrinsic errors in determining $AR_0$ it is still
interesting to compare the range of $AR_0$ implied by our data to $A$
measured from proper motions of stars.  Comparing our values of
$AR_0=89 - 110~{\rm km~s^{-1}}$ with the Hipparchos results for $A$
\citep{mignard00} shows that for $R_0=8.5$ the lower limit of our
range is consistent with the $A$ values derived for hot stars in the
disk (e.g.\ $A=10.9\pm0.8$ km s$^{-1}$ kpc$^{-1}$ for stars of type
A0-A5 with $|b|<30\arcdeg$), but even our upper limit is smaller than
the Hipparchos result for giant stars at all latitudes, which is $A\simeq
14.7\pm1$ km s$^{-1}$ kpc$^{-1}$.  Our upper limit value of $A$ assuming
$R_0=8.5$ kpc is also smaller than the conventionally accepted value
of $A=14.5$ km s$^{-1}$ kpc$^{-1}$ \citep{binney87}. This low value of
$A$ derived from the \HI\ terminal velocities compared with the
locally measured value based on stellar radial velocities and
distances supports the mounting evidence that the Galactic Center
distance is smaller than the 8.5 kpc assumed here
\citep{reid93,eisenhauer05}.

%---------------------------------------------------------------------
\subsection{Inner Galaxy Rotation Curve}
\label{subsec:rotcurve}
\begin{figure}
\centering
\includegraphics[width={\textwidth}]{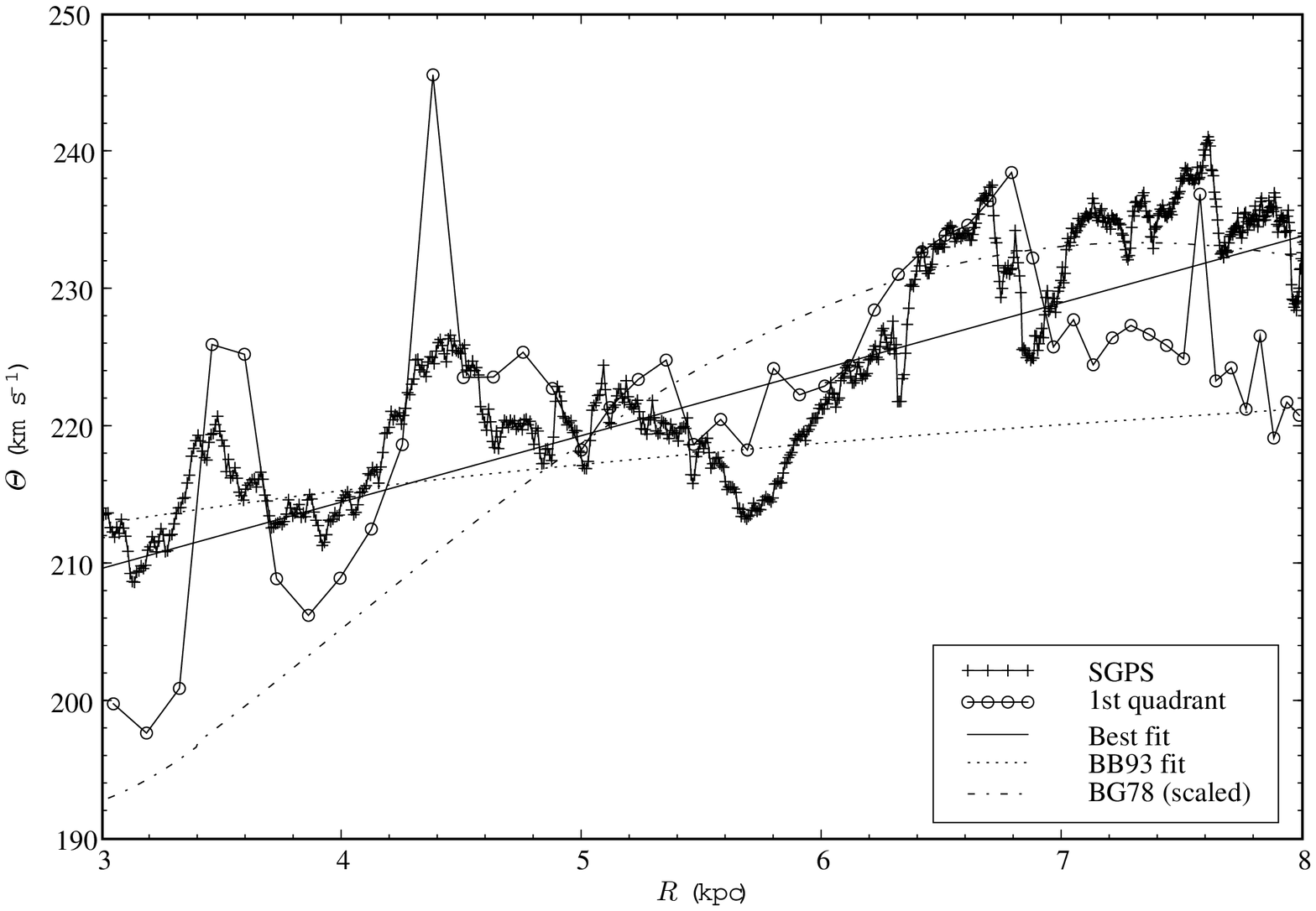}
\caption[]{Rotation curve for the fourth quadrant of the Milky Way
  interior to the Solar circle. The crosses are the SGPS data points
  and the open circles are the first quadrant values from
  \citet{fich89}.  Error bars are not shown on this plot to reduce
  confusion.  Estimates of the errors are included in the fitting and
  are $4.6$ for $l>0\arcdeg$ \citep{fich89},  $1$ \kms\ for
  $l<325\arcdeg$, $3$ \kms\ for   $l>332\fdg5$ and $10$ \kms\ for
  $327\fdg5 \leq l \leq 332\arcdeg$, as   described in the text.   The
  dotted line is the commonly used \citet{brand93} rotation curve
  derived for the outer Galaxy.  The dashed-dotted line is the
  \citet{burton78} polynomial fit to the first quadrant rotation
  curve.   This fit has been scaled from the original assumption of
  $R_0=10$ kpc and $\Theta_0=250$ \kms\ to $R_0=8.5$ kpc and
  $\Theta_0=230$ \kms.  We found it necessary to use $\Theta_0=230$
  \kms\ rather than $220$ \kms\ to place the peak near the observe
  peak velocities.  The solid line is the best fit to the ensemble of
  the first and fourth quadrant (SGPS) data over the range $3 \leq R
  \leq 8$ kpc given by $\Theta(R)/\Theta_0 = 0.186\pm0.008(R/R_0) +
  0.887\pm0.005 $, where $\Theta_0=220$ \kms\ and $R_0=8.5$ kpc.  
  \label{fig:rotcurve}}
\end{figure}

%---------------------------------------------------------------------
Using terminal velocities measured from the SGPS data and the circular
velocity relation that $\Theta (R) = |v_{LSR}| + \Theta_0 |\sin l|$ we
have calculated the rotation curve for $3 \leq R \leq 8$ kpc.  Figure
\ref{fig:rotcurve} shows the rotation curve for the fourth Galactic
quadrant derived from SGPS terminal velocity measurements.  Also
plotted on this figure are the rotation curve data points from the
first quadrant published by \citet{fich89}, which were derived from
\HI\ data presented by \citet{burton78}.  We include on
Figure~\ref{fig:rotcurve} two commonly used rotation curve fits, the
one derived for the outer Galaxy by \citet{brand93} and the one
derived for the first quadrant by \citet{burton78}.  Note that the
\citet{brand93} fit is almost identical to the \citet{fich89} rotation
curve fit, determined for the inner and the outer Galaxy.

The general trend of the SGPS rotation curve over the full range of
plotted radii is for increasing rotation velocity with increasing
galactocentric radius.  In addition, there is clearly a great deal of
dynamic structure along the locus of subcentral points.  The bump at
$R = 3.4$ kpc is attributed to the well-known three kiloparsec arm
\citep{vanwoerden57}. This feature is very well traced by the SGPS
terminal velocity data but it is clearly not in circular rotation and
should not fit by rotation curve models.  Other noticeable peaks and
troughs in the rotation curve have long been attributed to streaming
motions associated with the spiral arms of the Galaxy
\citep{shane66,yuan69,burton71}. At these locations the assumption of
circular rotation breaks down and the terminal velocity curve traces
not the overall rotation of the Galaxy, but the dynamical influence of
the spiral arms.

From $R\approx 3.9$ kpc the rotation curve rises rapidly from
$\Theta=210$ \kms\ to $225$ \kms\ at $R\approx 4.4$ kpc.  This rise is
mirrored in the first quadrant rotation curve, also shown in Fig.\
\ref{fig:rotcurve}, although for the first quadrant curve the rise is
shifted to larger radii by approximately 100 pc.  At $R \approx 5.7$
kpc there is a very sharp local minimum of $\Theta = 211$ \kms\
followed by a smooth rise towards a maximum of 240 \kms\ near
$R\approx 6.7$ kpc.  There is a clear difference between the first and
fourth quadrant rotation curves $R>7$ kpc. This difference has been
given as evidence of anomalous velocities associated with streaming
motion Sagittarius-Carina arm at $l \sim 295\arcdeg$
\citep{henderson82,grabelsky87}.  On the other hand, asymmetries
between the first and fourth quadrant curves have also been explained
by \citet{kerr62} and \citet{blitz91} as evidence for an outward
motion of the Local Standard of Rest orbit.  If there is an outward
motion of the LSR this would apply an offset to the derived rotation
curve of $-2\Pi_o\cos{l}$, where $\Pi_o$ is the magnitude of the
outward velocity, estimated to be $14$ \kms\ by \citet{blitz91}.  Our
current data, because they are restricted to the fourth quadrant, are
not sufficient to resolve whether there is an outward velocity of the
LSR so we omit this correction for now.

The SGPS data, because of their high angular resolution, allow us to
clearly resolve individual features in the rotation curve. Superposed
on the large scale features are a multitude of small features with
velocity amplitudes $\sim 2-5$ \kms, over length scales of 100 to 200
pc. One such example lies at $\sim 8$ kpc and can be attributed to the
Carina supershell \citep{dawson06}, which has carved a hole out of the
disk and disturbed the kinematics near the terminal
velocity. Presumably some of the other small scale features can be
attributed to discrete objects, but another fraction must represent
the random cloud-to-cloud motions near the subcentral point.

\subsubsection{A Fit to the Rotation Curve for $3 \leq R \leq 8$ kpc}
\label{subsubsec:fit}
As seen in Figure~\ref{fig:rotcurve}, neither of the two commonly used
rotation curve formulae are good fits to the data.  The
\citet{brand93} rotation curve, which is often used for kinematic
distance estimates, is not a particularly good representation of both
the first and fourth quadrants.  That this curve is a poor fit for the
inner Galaxy ($R<R_0$) is not surprising as it was derived to fit the
outer Galaxy.  When plotted over the inner Galaxy it is clear that the
\citet{brand93} fit is much flatter than the data. Similarly, the
\citet{burton78} fit, although a reasonable fit for the first
quadrant, is much too steep for the fourth quadrant data.  Other
rotation curves have been derived specifically for the inner Galaxy,
such as \citet{clemens85}, but this high order polynomial fit was
derived to fit data from the locus of subcentral points in the first
quadrant.  The curve includes fits to the large scale velocity
deviations known to be associated with spiral arms and is therefore of
limited applicability to other parts of the inner Galaxy.

Here we use both the first and fourth quadrant data together to derive
a new fit to the rotation curve appropriate for $3~{\rm kpc}<R <
8~{\rm kpc}$ at all longitudes.  Following \citet{fich89} and
\citet{brand93} we fit the data with a variety of functional forms,
including first and second-order polynomials and the power law $\Theta
/\Theta_0 = c_1 (R/R_0)^{c_2} + c_3$, where $c_1$, $c_2$, and $c_3$
are the free parameters to be determined by the fit.  The fitting was
done using a non-linear least-squares algorithm on the combination of
the first quadrant and SGPS data, including the standard deviations of
each.  The standard deviation for the first quadrant velocities is
$\sigma_v=4.5$ \kms\ \citep{fich89,burton78}.  The SGPS standard
deviations were derived from the repeatability of the $v_{LSR}$
determination as discussed in \S \ref{subsec:errfunc}.  We let
$\sigma_v=1$ \kms\ for $l<325\arcdeg$ and $\sigma_v=3$ \kms\ for
$l>332\fdg5$ and $\sigma_v=10$ \kms\ for $327\fdg5 \leq l \leq
330\arcdeg$.  All data were additionally weighted according to
$\sin{l}$ to compensate for the uneven sampling in this domain.
Finally, the first quadrant weights were adjusted by the ratio of the
number of points in the first and fourth quadrant datasets to
compensate for the fact that the SGPS data are much more densely
sampled than the first quadrant data.  We found that the best fit for
the range of radii used here, which minimised the absolute
chi-squared, was a simple linear function, where
\begin{equation}
\Theta(R)/\Theta_0 = (0.186\pm0.008) (R/R_0) + (0.887\pm0.005) 
\label{eq:rotfit}
\end{equation}
This fit is shown in Figure~\ref{fig:rotcurve}.  We emphasize that
this fit is specifically intended for the region $3~{\rm kpc}<R <
8~{\rm kpc}$ and is not a good fit to outer Galaxy rotation.  This
fit, however, is generalised for applicability to the inner Galaxy in
both the first and fourth Galactic quadrants.  Although we have
assumed $\Theta_0=220$ \kms\ in Figure~\ref{fig:rotcurve}, the fit is
independent of the assumed values for $\Theta_0$ and $R_0$.

\subsubsection{Velocity Residuals}
\label{susubsec:resid}
From Figure~\ref{fig:rotcurve} it is clear that there are significant
motions that are not fit by this simple rotation curve and the
magnitude of these motions should be considered when using this fit
for kinematic distances.  These residuals give an estimate for the
magnitude of streaming motions in the inner Galaxy.
Figure~\ref{fig:resid} shows the residuals for the SGPS and 1st
quadrant rotation curves minus the linear fit described above.  The
peak-to-peak variations in the SGPS residuals using the new fit are
$10$ \kms\ and they have a standard deviation of $5.1$ \kms.  The
first quadrant residuals have a larger peak-to-peak variation but this
should be confirmed by better sampled data.  For comparison, the
standard deviation of the SGPS residuals from the \citet{brand93} fit
are $6.6$ \kms.  The magnitude of these residuals does not differ
significantly from those estimated by \citet{shane66,burton71} and
others, but is perhaps the most well determined estimation available.

That the positions of the bumps in the rotation curve relate to spiral
structure has been long known and successfully demonstrated by
applications of density wave theory to \HI\ observations of the Milky
Way \citep[e.g.][]{yuan69,burton71,roberts69}, and other galaxies
\citep[e.g.][]{rots75,visser80}.  Indeed, these bumps are the
strongest indicators we have of the locations of the spiral arms.  The
problem, however, has been that the linear and non-linear versions of
the density wave theory predict different relationships between the
gravitational centre of a spiral arm and the associated velocity peak.
Past data from the Milky Way has not allowed us to distinguish between
these two theories, but with the abundance of high resolution \HI\
\citep{mcgriff05,stil06}, CO \citep{luna06}, and infrared data
\citep{benjamin03} for the inner Milky Way now available it may be
time to revisit this topic.  This will be the topic of a future paper
in this series.  

%----------------------------------------------------------
\section{Random Motions Determined From The Terminal Velocity Shape}
\label{sec:random}
%----------------------------------------------------------
In addition to providing the offset velocity, $v_o$, the shape of the
terminal velocity profile provides information about random gas
motions in the ISM.  The subcentral point is a useful location for
measuring these quantities because velocity crowding forces us to
observe a large ensemble of \HI\ ``clouds''.  When we measure the
velocity dispersion of \HI\ near the subcentral point we measure a
mixture of the microscopic (thermal) and macroscopic (the so-called
cloud-to-cloud velocity dispersion) motions.  The terminal profile
shape is therefore one of the best probes of cloud-to-cloud velocity
dispersion in the Galaxy.  Our error function fits return values for
the velocity widths of the error functions, which are related to the
velocity dispersions of the gas as $\Delta v_i=\sqrt{2}\sigma_v$.
Using the SGPS data beyond the subcentral point we have studied the
velocity widths both for the dataset as a whole and as a function of
Galactic radius.

In Table ~\ref{tab:errfit} we give values for the fitted parameters
for two and three component error function fits to the longitude
averaged terminal velocity profile.  The fitted parameters, $a_i$,
$v_o$ and $\Delta v_i$, are those defined in
Equation~\ref{eq:errfunc}.  The two component fit parameters are given on
the first line of Table \ref{tab:errfit} and the three component fit
parameters are on the second line.  The quoted errors are the
variation in the parameters that double the reduced $\chi^2$ of the
fit.  Note that these errors do not account for co-variance of the
fitting parameters.  As mentioned in \S \ref{subsec:errfunc} and shown
in Figure~\ref{fig:errfav} three components produce a significantly
better fit to the data, reducing the reduced chi-squared of the fit
from over 100 for two components to $\sim 4$ for three components.

The velocity widths of the two component fit are similar to the
\citet{kulkarni85} observations of \HI\ emission towards $l=30\arcdeg$
and $l=180\arcdeg$, which revealed a narrow component with a velocity
dispersion of $\sigma_v =6 - 8$ \kms\ and a second component with a
velocity dispersion of $15 - 35$ \kms.  \citet{kulkarni85} found that
the fits to the broad component were not unique and that values
between 15 \kms\ and 35 \kms\ fit the data equally well.  The \HI\
spectra that \citet{kulkarni85} used were from \citet{weaver73}, with
an rms noise of $\sigma_T =0.38$ K, whereas the SGPS longitude
averaged spectrum probes the tail of the \HI\ profile to $\sigma_T\sim
50$ mK.  With the better sensitivity of the longitude averaged SGPS
spectrum we can clearly see that the data are better described with
three velocity components.  Compared to the two component fit, the
two narrowest components of the three component fit are smaller.  We
see that in the three component fit the amplitudes of the first two
components are similar and the amplitude of the third component is
$\sim 10$\% of the first.  

That the terminal velocity profile requires three components is
perhaps not surprising.  \HI\ observed in emission is a mixture of
warm ($T\sim 6000$ K) and cool ($T=50 -100$ K) kinetic temperatures.
If we ignore the effects of the velocity gradient, the observed
velocity dispersion, $\sigma_v$, represents a combination of the
thermal velocity width, $\sigma_{th}$, and the random velocity of the
cloud ensemble (cloud-to-cloud motions), $\sigma_{cc}$, such that the
total velocity width is $\Delta v = \sqrt{2}\sigma_v = \sqrt{2 \,
  (\sigma_{th}^2 + \sigma_{cc}^2)}$.  If we assume thermal velocity
dispersions for the cold and warm gas of $\sim 1$ \kms\ and $\sim 7$
\kms, respectively, then the widths of the two narrow components give
cloud-to-cloud velocity dispersions of $\sim 4.4$ \kms\ and $\sim 5.2$
\kms, which are identical to within the errors.  This suggests that
the cold and warm \HI\ share the same bulk motions.

\subsection{Velocity Dispersion Variations with Galactocentric Radius}
\citet{kulkarni85} suggested that the wide velocity dispersion might
be related to shocked clouds.  For this reason they chose to examine a
profile at $l=30\arcdeg$ where the line-of-sight runs tangent to the
$R=5$ kpc annulus where supernova remnants show a peak.  Additionally,
Koo and collaborators \citep[i.e.][]{koo04} have suggested that
extended forbidden-velocity line wings may be associated with
individual supernova remnants, whose shocks drive \HI\ to high
velocities.  Given these suggestions it is interesting to explore
whether there may be a significant longitude dependence to the
velocity dispersion.

Using the full range of the SGPS data we explore whether the velocity
widths of the three error functions change with position in the
Galaxy.  For this analysis we include data in the range $285\arcdeg
\leq l \leq 340\arcdeg$, which excludes data from the inner 3 kpc of
the Galaxy where the ISM environment is different from the rest of the
disk.  We have also excluded data for $l<285\arcdeg$ because at these
longitudes the velocity gradient, $dv/dr$, is near zero for a
particularly long path length, leading to saturation in the \HI\
profiles.  Over the longitude range $285\arcdeg \leq l \leq
340\arcdeg$ there are $\sim 1520$ independent samples.  However, to
explore the broad velocity component we are interested in brightness
temperatures extending as low as $\sim 250$ mK, so we must average the
data into longitude bins of $\sim 1\fdg4$ to achieve sufficient
signal-to-noise. Note that this analysis necessarily precludes us from
finding possible missing supernova remnants by their
forbidden-velocity line-wings because we average over large spatial
ranges.  For each longitude bin we have fit the terminal velocity
curve with a sum of three error functions, as defined in
Equation~\ref{eq:errfunc}.  The difference between three and two error
functions is important here as a fit with only two components produces
a consistently wider first component than does the three component
fit.

In Figure \ref{fig:erflong} we plot the widths of the three error
functions, $\Delta v_1(R)$, $\Delta v_2(R)$, and $\Delta v_3(R)$ as a
functions of galactocentric radius.  The error bars are estimated as
the variation in the parameter that results in a doubling of the
reduced chi-squared.  In this plot the three different symbols
correspond to the three different components.  The mean values of the
three widths are: $6.5\pm2.1$ \kms, $12.8\pm3.2$ \kms\ and $25.5 \pm
7.9$ \kms, which agree well with the values found for the fit to the
data averaged over the entire longitude range.  We have fit both of
the narrowest components as functions of radius, shown in
Figure~\ref{fig:erflong}.  The slope of the narrowest component is
consistent with zero: $\Delta v_1(R) = (6\pm1) +(0.2\pm0.2) R$ \kms.
There is only a very slight slope to the second component, which
goes as $\Delta v_2(R) = (13\pm1) - (0.3\pm0.2)R$ \kms. The widest
component does not show a linear trend but shows a significant
increase at small Galactic radii.  This increase is accompanied
with increasing errors because at large longitudes the tail of the
\HI\ emission extends to the edge of the SGPS bandwidth, making it
difficult to fit the long tail of the \HI\ profile.  Despite the
larger errors, the increase appears real.

\citet{burton71} also examined the velocity dispersion versus Galactic
radius, fitting his single component velocity dispersion with a
function of the form $\sigma(R) = 9-0.4R$ \kms.  Our fit for the narrowest
component is consistent with no gradient in the velocity dispersion
across the disk and the middle component shows a shallower slope than
Burton's result.  At first our result appears inconsistent with
observations of velocity dispersion across the disk of face-on
external galaxies, which, like \citet{burton71} show that inside
the stellar disk the narrow component velocity dispersion increases
with decreasing radius \citep{dickey90b}.  The resolution of this
apparent conflict, however, may lie in the fact that both studies were
based on single component fits, which will produce wide velocity
dispersions if the profiles would be better fit by two or three
Gaussian components.  From our work it seems that the narrow
components are effectively constant with radius while the broad
component increases with decreasing radius.

In addition to the large-scale trends discussed above, all three
components show some small-scale features.  At $R\sim 5$ kpc the wide
component increases by a factor of $\sim 2$ over a width of $\sim 500$
pc.  This increase is also reflected in increases in both of the
narrow components. We see that at $R\sim 6.3$ kpc and $R\sim 6.8$ kpc
the fit effectively reduces to a two component model.  It may be
noteworthy that the narrowest two components track each other very
well with Galactic radius, with all fluctuations observed in both
components.  This observation supports our suggestion in the previous
section that the warm and cold \HI\ share the same bulk motions, which
may be either large-scale streaming motions or cloud-to-cloud
motions. The separation between the two widths is simply the
difference in the thermal motions, not the macroscopic motions which
appear to effect warm and cold gas equally.

Some observations of face-on galaxies suggest that there may be some
correlation between velocity dispersion and spiral structure
\citep[e.g.][]{braun97}. Similarly, the original supposition of
\citet{kulkarni85} that led them to look for ``fast'' clouds at
$l=30\arcdeg$ was that there may be correlation between fast clouds
and areas rich with supernovae, such as spiral arms.  At first glance
Figure~\ref{fig:erflong} seems to support this latter supposition
because all velocity components increase near $R=5$ kpc. It is
tempting to look deeper into our fitted velocity dispersions for
correlation with spiral structure but \citet{burton71} reminds us that
at the subcentral point the measured velocity dispersion can be
related to the path length contributing to the velocities near the
subcentral point, in other words the inverse of the velocity gradient,
$|dv/dr|$.  In the presence of streaming motions $|dv/dr|$ is altered
from the value predicted by circular rotation.  An increase in
velocity dispersion is therefore not necessarily indicative of an
increase in the cloud-to-cloud dispersion, but could simply indicate
an increase in the path length contributing to the emission near the
terminal velocity.  Burton showed that for linear density wave theory
streaming motions should increase the path length near the terminal
velocity just {\em outside} the spiral arms and suggested that there
was an indicative increase in his measured dispersions.  However, the
\citet{burton71} analysis was hampered by single Gaussian fits, which,
as he pointed out, were extremely difficult at longitudes where the
velocity component became so broad with a flat top that it was better
fit by two separate components.  Fitting error functions to the tail
of the emission beyond the terminal velocity provides some protection
against this problem but there is no doubt that in the presence of
streaming motions the measured velocity dispersion will change from
that measured given purely circular rotation.  Unfortunately, once
again the {\em location} of increases relative to spiral structure
depends on the density wave theory applied, linear or
non-linear. Without knowing the gravitational positions of the spiral
arms it is impossible to say whether the increases in velocity width
that we observe in all components at $R\sim 5$ kpc and in the largest
component at $R<4$ kpc are indicative of a change in $dv/dr$ or a
change in the cloud-to-cloud velocity dispersion.

The velocity gradient also plays a role in the {\em mixture} of
gas-phase components measured beyond the terminal velocity.  In the
simplest case this can be understood if we consider that for emission
beyond the subcentral point and in a given velocity interval, $\delta v$,
the path length contributing to the emission is $\delta r= \delta v
(dv/dr)^{-1}$.  We can therefore see that for gas with a narrow
velocity dispersion only gas very near to the subcentral point contributes
to the emission beyond the terminal velocity, whereas gas with a large
velocity dispersion can be distributed over a large path length near the
subcentral point and still contribute to the emission beyond the
terminal velocity.  Precisely at the subcentral point the velocity
gradient, $\left|dv/dr\right|=0$ so that it is the second derivative
of velocity with distance, $d^2v/dr^2$, that determines the distance
along the line of sight that contributes to an interval of velocity
near the subcentral point as $\delta r = \sqrt{2 \ \delta v \left|
    d^2v/dr^2 \right|^{-1}}$, i.e.\ the second term of a Taylor
expansion of $v(r)$ for $r\sim r_0$.  The larger the second
derivative, the smaller the line of sight contributing to a given
spectral interval.  Even in pure circular rotation the second
derivative of velocity near the subcentral point changes as a function
of longitude so we expect that the proportion of phases represented
beyond the terminal velocity will change with longitude.  As discussed
above, the velocity gradient is also sensitive to streaming motions so
understanding the mixture of phases with radius is an extremely
complicated issue.  We defer that discussion to a future paper, which
will explore the spiral structure more fully.

%----------------------------------------------------------
\section{Conclusions}
\label{sec:conclusions}
\begin{figure}
\centering
\includegraphics[width={\textwidth}]{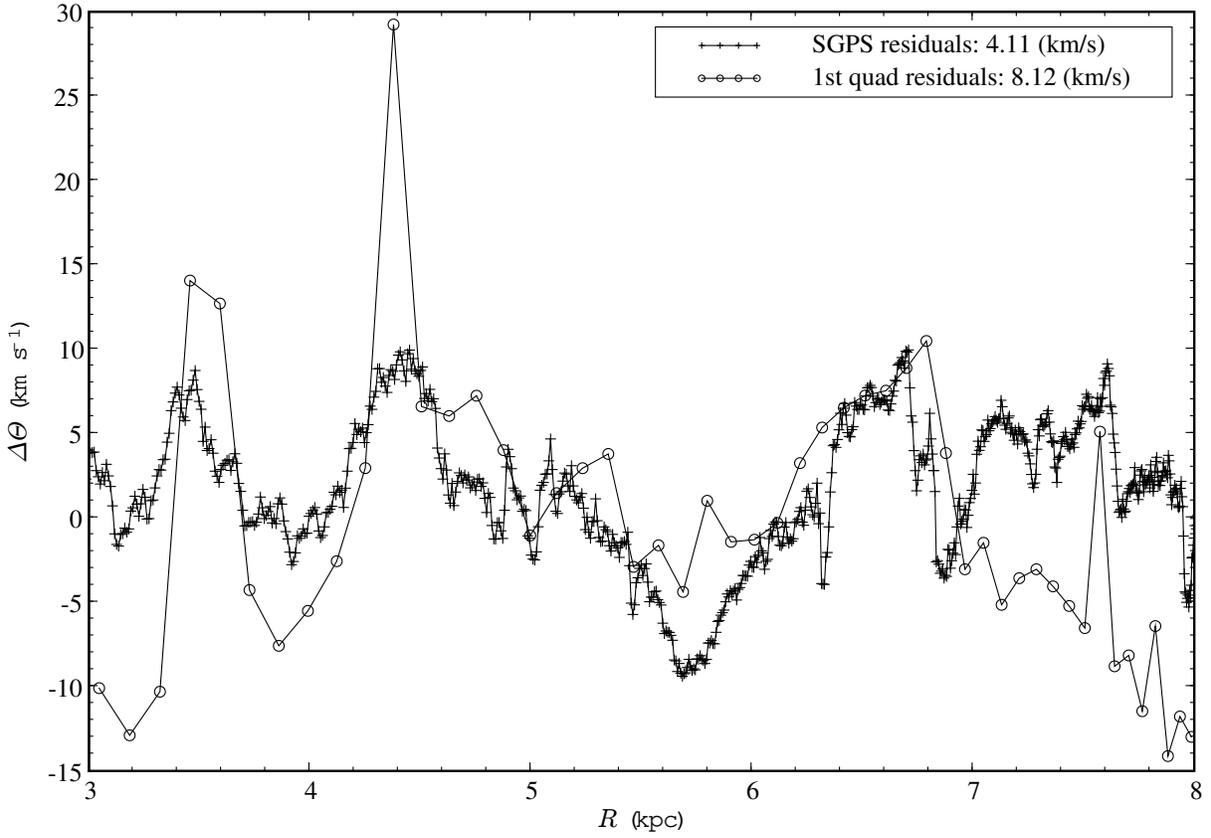}
\caption[]{Residuals from the rotation curve fit presented in Figure
  \ref{fig:rotcurve}.  The standard deviation for the SGPS residuals is
  4.11 \kms, whereas the standard deviation for residuals from the
  \citet{brand93} curve was 6.6 \kms.  
\label{fig:resid}}
\end{figure}

%----------------------------------------------------------
For the $\lambda$21-cm pioneers who first measured the Milky Way
rotation curve 50 years ago, the objective was to determine the radial
variation of the gravitational potential, and so to measure the
distribution of gravitational mass.  In this they were quite
successful \citep{kwee54,schmidt56} and only relatively minor
adjustments have been made since by breaking the Galaxy down into its
disk and spheroidal components \citep{caldwell81}.  Refinements such
as those discussed in this paper do not alter the overall result for
the mass of the Galaxy.  The significance of the high angular
resolution and high precision measurement techniques developed here
have more to do with departures from circular rotation than with the
overall mass versus radius function.  We have fit a new rotation curve
to the region $3 \leq R \leq 8$ kpc using our data from the fourth
quadrant together with a previously published rotation curve for the
first quadrant.  This generalised fit attempts to overcome problems
related to quadrant specific fits and is applicable for all of the
inner Galaxy. Using this fit we determine the velocity residuals for
the fourth quadrant confirming with greater precision that the
streaming motions have amplitudes of $\sim 10$ \kms.  Although the
wiggles evident on Figure~\ref{fig:resid} have long been interpreted
in terms of spiral structure their precise location with respect to
the gravitational centre of spiral arms is theory dependent. We have
future work planned to combine high precision \HI\ and CO surveys with
observations of star-forming regions to better quantify the spiral
structure in the inner Galaxy.

\begin{figure}
\centering
\includegraphics[width={\textwidth}]{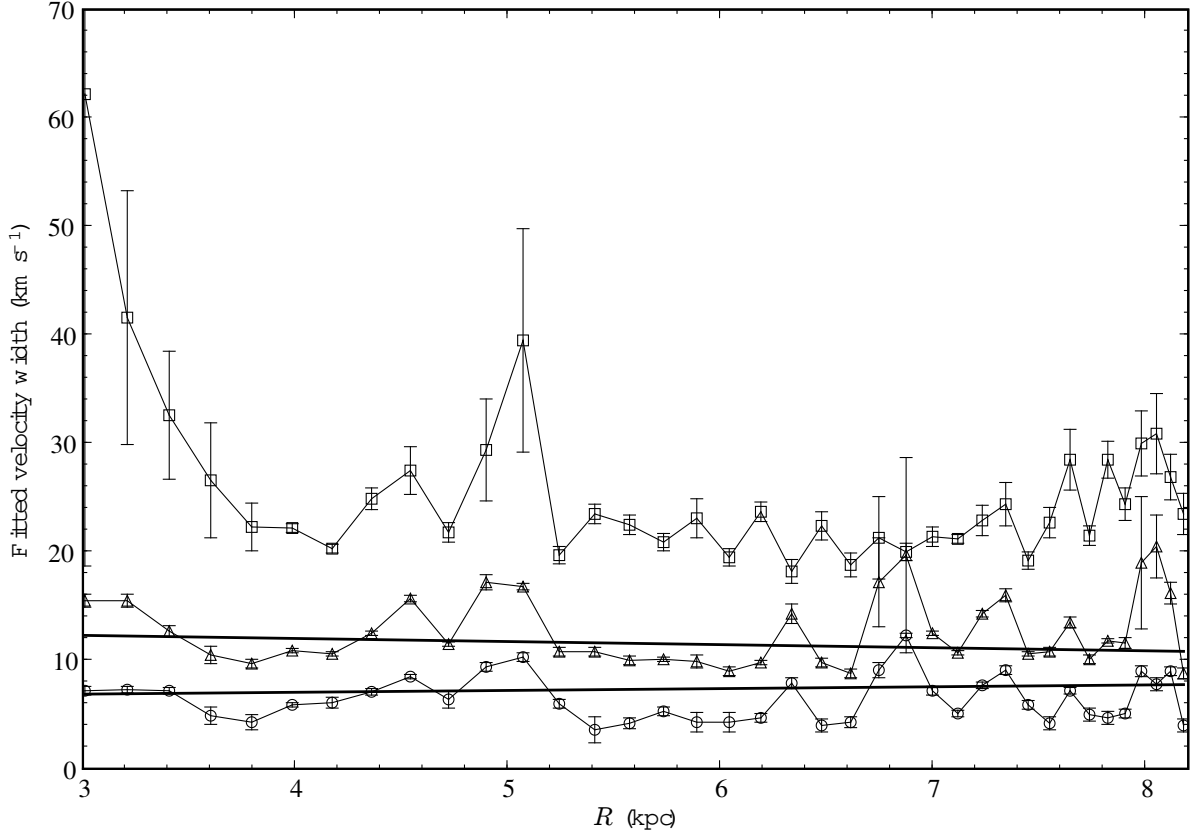}
\caption[]{Widths of the three velocity components fitted to the
  terminal velocity profile as a function of Galactic longitude.
  These data are binned by $1\fdg4$ of Galactic longitude.  The
  plotting symbols for the narrow, middle and wide components are
  circles, triangles and boxes, respectively.  The error bars are the
  amount of variation in a parameter required to double the
  chi-squared of the fit.  The narrowest components are fit with
  linear functions of the form $\Delta v_1(R) = (6\pm1) +
  (0.2\pm0.2)R$ \kms and $\Delta v_2(R) = (13\pm1) -(0.3\pm0.2)R$
  \kms.    
  \label{fig:erflong}}
\end{figure}

In addition to large scale variations in the magnitude and direction
of the streamlines of the interstellar gas as it rotates around the
Galactic Center, some of the small scale fluctuations measured here
may be due to non-gravitational perturbations that change the shape of
the line profile beyond the terminal velocity.  In \S
\ref{sec:termvel} and \S \ref{sec:random} we carefully discussed the
velocity offset derived from error function fitting to the profile
shape beyond the terminal velocity, because the random motions
reflected in this shape must be separated from large scale variations
in the rotation velocity.  Thus in order to trust the interpretation
of the wiggles on Figure~\ref{fig:resid} as being due to spiral arms,
we must first eliminate wiggles in $v_o$ similar to those in $\Delta
v_1$, $\Delta v_2$, and $\Delta v_3$ on Figure \ref{fig:erflong}. 

\begin{figure}
\centering
\includegraphics[angle=-90,width={\textwidth}]{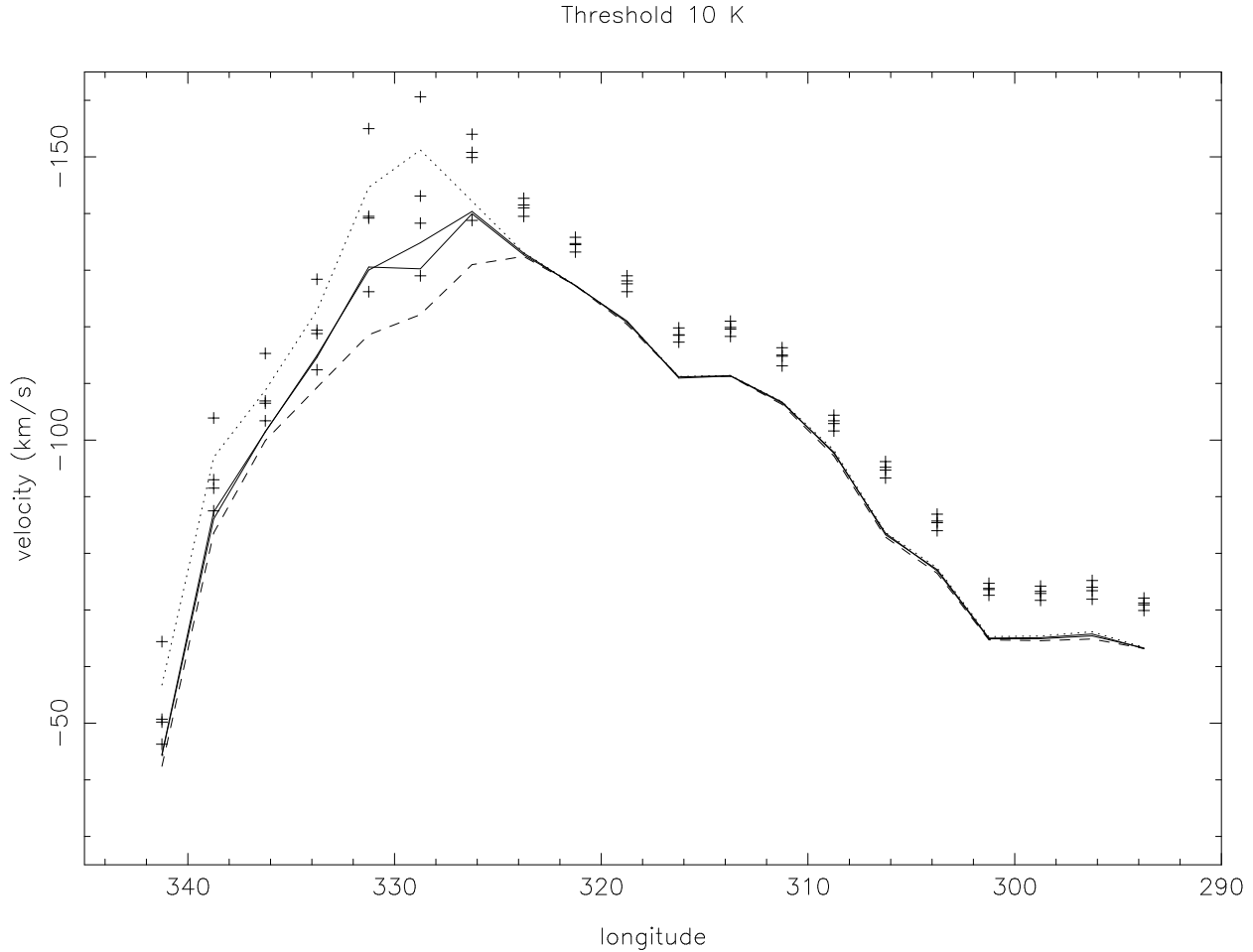}
\caption[]{The effect of correcting the threshold crossing velocity by
  $v_o$, the velocity offset from the error function fits.  The four
  crosses at each longitude are the average values of $v_t$ in each
  longitude interval ($2\fdg5$), for the four statistics (mean,
  median, 25th and 75th percentile) used to find the threshold
  crossings.  The 25th percentile gives consistently lower negative
  velocity values, while the 75th percentile gives higher values,
  while the mean and median give values of $v_t$ that are generally
  quite close together.  But after adding the corresponding $v_o$ all
  the values agree much better, as indicated by the curves that join
  values of $v_{LSR}$ for each longitude bin.  The dotted curve is for
  the 75th percentile, the dashed curve for the 25th percentile, and
  the two solid curves are for the mean and median.  All four agree to
  better than 0.5 \kms\ for longitudes less than about $l=325\arcdeg$, as
  discussed in the text.
\label{fig:plotrot}}
\end{figure}

\begin{figure}
\centering
\includegraphics[angle=-90,width={\textwidth}]{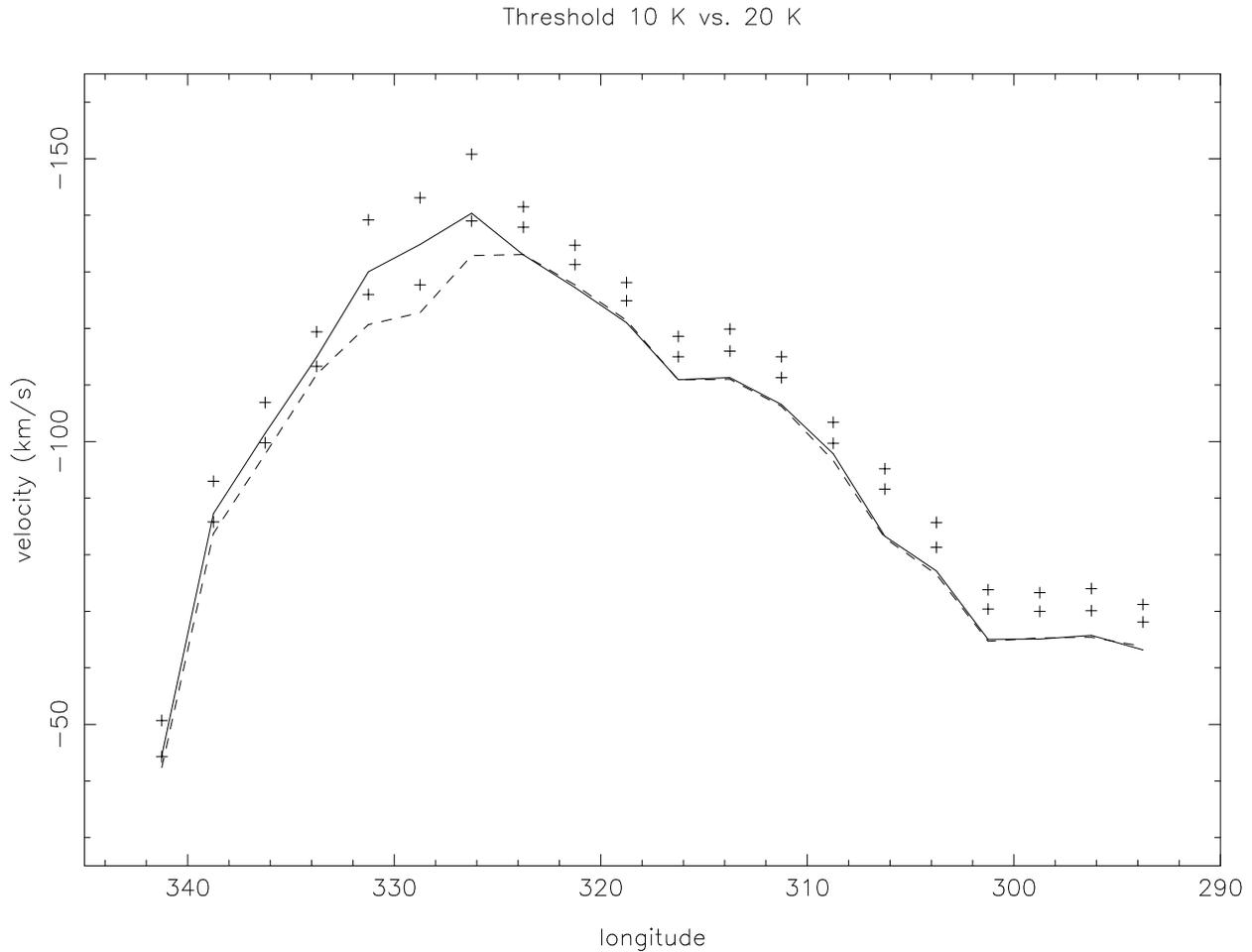}
\caption[]{The difference between choosing thresholds of 10 K and 20 K in
brightness temperature.  As on Figure \ref{fig:plotrot} the crosses
show the threshold crossing velocities, this time for the two
thresholds both using the mean value (average over latitude) as the
statistic.  The higher negative velocities correspond to the lower
threshold, as  shown on fig.~\ref{fig:plotrot}.  After applying the
correction $v_o$ found from erfc fitting, the two thresholds give the
same result to less than 0.5 \kms\ for longitudes less than $325\arcdeg$,
as discussed in the text.  This demonstrates that the final function
$v_{LSR}(l)$ is independent of the arbitrary choice of threshold over
most of the longitude range of the survey.
\label{fig:plotrot2}}
\end{figure}

We have presented a successful technique for measuring the terminal
velocity of \HI\ in the inner Galaxy.  As shown in Figures
\ref{fig:plotrot} and \ref{fig:plotrot2}, the technique of fitting
complementary error functions to the \HI\ terminal velocity is quite
robust, giving results that are consistent to less than the width of
one velocity channel over much of the longitude range of the
survey. Our technique is effective at overcoming the biases that might
be introduced by determining the terminal velocity using brightness
temperature threshold methods. Our fits show that \HI\ terminal
velocities estimated based on a brightness temperature threshold
without any correction will over-estimate the terminal velocity by
$5-10$ \kms.  Although the CO velocity dispersion is smaller than for
\HI, terminal velocities estimated from CO emission using a threshold
technique are also not immune from over-estimation by amounts
comparable to molecular cloud velocity dispersion.  Comparison of the
\HI\ and \citet{luna06} $^{12}$CO terminal velocities has shown that
the CO velocities have more scatter than those presented here from
\HI.  We attribute some of this difference to the different volume
filling factors of \HI\ and CO and suggest that \HI\ measurements of
the terminal velocity may be more robust than CO because they detect a
larger ensemble of clouds.

The technique of fitting the \HI\ terminal velocity profile has the
added benefit that it returns information about the velocity width of
the \HI\ beyond the terminal velocity.  The fitted velocity width
combines information about the thermal width of the gas, the inherent
velocity dispersion as caused by cloud-to-cloud motions, and changes
in the velocity gradient, which may be caused by streaming motions
associated with spiral arms.  Unravelling these effects at specific
positions within the inner Galaxy is complicated and something we hope
to address in a future paper.  Averaging over all longitudes to
mitigate against local variations in the velocity gradient we have
shown that the \HI\ terminal velocity shape is best fit by the
combination of three error functions, with widths of $6$ \kms, $12$
\kms, and $26$ \kms.  Examining the three velocity widths as a
function of Galactic radius we found that the widths of the narrower
two components track each other very well at all Galactic radii and
show no appreciable trend with radius. We suggest that the agreement
between the narrow components is evidence that the two thermal phases
share the same bulk motions as well as variations in the velocity
gradient.  Unlike the two narrow components, the third, widest,
component appears to increase with decreasing Galactic radius for
$R<4$ kpc.  It is unclear at this time whether this increase reflects
a true increase in the cloud-to-cloud motions or streaming motions.

%--------------------------------------------------------- 
\acknowledgements This research was supported in part by NSF grants
AST-9732695 and AST-0307603 to the University of Minnesota.  We
greatly appreciate the involvement and advice of B.\ M.\ Gaensler and
A.\ J.\ Green with the original SGPS observations and all subsequent
analysis.  NMMc-G acknowledges D.\ McConnell for suggestions related
the automated error function fitting of \HI\ profiles and A.\ Luna for
providing the CO terminal velocities in tabulated form.  Both NMMc-G
and JMD acknowledge comments and suggestions by W.\ B.\ Burton on a
previous version of this manuscript.  We would particularly like to
thank E.\ Levine and the referee, F.\ J.\ Lockman, for excellent
questions and suggestions.
%--------------------------------------------
% Bibliography
%--------------------------------------------

%\bibliographystyle{apj} 
%\bibliography{references} %~naomi/tex/references.bib, bibtex file 

\small 
\begin{thebibliography}{49}
\expandafter\ifx\csname natexlab\endcsname\relax\def\natexlab#1{#1}\fi

\bibitem[{Alvarez {et~al.}(1990)Alvarez, May, \& Bronfman}]{alvarez90}
Alvarez, H., May, J., \& Bronfman, L. 1990, \apj, 348, 495

\bibitem[{{Anantharamaiah} {et~al.}(1984){Anantharamaiah}, {Radhakrishnan}, \&
  {Shaver}}]{anantharamaiah84}
{Anantharamaiah}, K.~R., {Radhakrishnan}, V., \& {Shaver}, P.~A. 1984, \aap,
  138, 131

\bibitem[{{Benjamin} {et~al.}(2003){Benjamin}, {Churchwell}, {Babler}, {Bania},
  {Clemens}, {Cohen}, {Dickey}, {Indebetouw}, {Jackson}, {Kobulnicky},
  {Lazarian}, {Marston}, {Mathis}, {Meade}, {Seager}, {Stolovy}, {Watson},
  {Whitney}, {Wolff}, \& {Wolfire}}]{benjamin03}
{Benjamin}, R.~A., {Churchwell}, E., {Babler}, B.~L., {Bania}, T.~M.,
  {Clemens}, D.~P., {Cohen}, M., {Dickey}, J.~M., {Indebetouw}, R., {Jackson},
  J.~M., {Kobulnicky}, H.~A., {Lazarian}, A., {Marston}, A.~P., {Mathis},
  J.~S., {Meade}, M.~R., {Seager}, S., {Stolovy}, S.~R., {Watson}, C.,
  {Whitney}, B.~A., {Wolff}, M.~J., \& {Wolfire}, M.~G. 2003, \pasp, 115, 953

\bibitem[{Binney \& Tremaine(1987)}]{binney87}
Binney, J. \& Tremaine, S. 1987, Galactic Dynamics (Princeton: Princeton
  University Press)

\bibitem[{{Blitz} \& {Spergel}(1991)}]{blitz91}
{Blitz}, L. \& {Spergel}, D.~N. 1991, \apj, 370, 205

\bibitem[{{Brand} \& {Blitz}(1993)}]{brand93}
{Brand}, J. \& {Blitz}, L. 1993, \aap, 275, 67

\bibitem[{{Braun}(1997)}]{braun97}
{Braun}, R. 1997, \apj, 484, 637

\bibitem[{{Burton}(1971)}]{burton71}
{Burton}, W.~B. 1971, \aap, 10, 76

\bibitem[{{Burton}(1976)}]{burton76}
---. 1976, \araa, 14, 275

\bibitem[{{Burton} \& {Gordon}(1978)}]{burton78}
{Burton}, W.~B. \& {Gordon}, M.~A. 1978, \aap, 63, 7

\bibitem[{{Burton} \& {Liszt}(1993)}]{burton93}
{Burton}, W.~B. \& {Liszt}, H.~S. 1993, \aap, 274, 765

\bibitem[{{Caldwell} \& {Ostriker}(1981)}]{caldwell81}
{Caldwell}, J.~A.~R. \& {Ostriker}, J.~P. 1981, \apj, 251, 61

\bibitem[{{Celnik} {et~al.}(1979){Celnik}, {Rohlfs}, \&
  {Braunsfurth}}]{celnik79}
{Celnik}, W., {Rohlfs}, K., \& {Braunsfurth}, E. 1979, \aap, 76, 24

\bibitem[{{Clemens}(1985)}]{clemens85}
{Clemens}, D.~P. 1985, \apj, 295, 422

\bibitem[{{Dawson} {et~al.}(2006){Dawson}, {Kawamura}, \& {Fukui}}]{dawson06}
{Dawson}, J., {Kawamura}, A., \& {Fukui}, Y. 2006, in IAU Symposium

\bibitem[{{Dickey} {et~al.}(1990){Dickey}, {Hanson}, \& {Helou}}]{dickey90b}
{Dickey}, J.~M., {Hanson}, M.~M., \& {Helou}, G. 1990, \apj, 352, 522

\bibitem[{{Dickey} \& {Lockman}(1990)}]{dickey90}
{Dickey}, J.~M. \& {Lockman}, F.~J. 1990, \araa, 28, 215

\bibitem[{{Eisenhauer} {et~al.}(2005){Eisenhauer}, {Genzel}, {Alexander},
  {Abuter}, {Paumard}, {Ott}, {Gilbert}, {Gillessen}, {Horrobin}, {Trippe},
  {Bonnet}, {Dumas}, {Hubin}, {Kaufer}, {Kissler-Patig}, {Monnet},
  {Str{\"o}bele}, {Szeifert}, {Eckart}, {Sch{\"o}del}, \&
  {Zucker}}]{eisenhauer05}
{Eisenhauer}, F., {Genzel}, R., {Alexander}, T., {Abuter}, R., {Paumard}, T.,
  {Ott}, T., {Gilbert}, A., {Gillessen}, S., {Horrobin}, M., {Trippe}, S.,
  {Bonnet}, H., {Dumas}, C., {Hubin}, N., {Kaufer}, A., {Kissler-Patig}, M.,
  {Monnet}, G., {Str{\"o}bele}, S., {Szeifert}, T., {Eckart}, A.,
  {Sch{\"o}del}, R., \& {Zucker}, S. 2005, \apj, 628, 246

\bibitem[{{Ferri\`{e}re} {et~al.}(2007){Ferri\`{e}re}, {Gillard}, \&
  {Jean}}]{ferriere07}
{Ferri\`{e}re}, K.~M., {Gillard}, W., \& {Jean}, P. 2007, \aap, in press

\bibitem[{{Fich} {et~al.}(1989){Fich}, {Blitz}, \& {Stark}}]{fich89}
{Fich}, M., {Blitz}, L., \& {Stark}, A.~A. 1989, \apj, 342, 272

\bibitem[{{Grabelsky} {et~al.}(1987){Grabelsky}, {Cohen}, {Bronfman},
  {Thaddeus}, \& {May}}]{grabelsky87}
{Grabelsky}, D.~A., {Cohen}, R.~S., {Bronfman}, L., {Thaddeus}, P., \& {May},
  J. 1987, \apj, 315, 122

\bibitem[{Gunn {et~al.}(1979)Gunn, Knapp, \& Tremaine}]{gunn79}
Gunn, J.~E., Knapp, G.~R., \& Tremaine, S.~D. 1979, \aj, 84, 1181

\bibitem[{{Haverkorn} {et~al.}(2006){Haverkorn}, {Gaensler},
  {McClure-Griffiths}, {Dickey}, \& {Green}}]{haverkorn06}
{Haverkorn}, M., {Gaensler}, B.~M., {McClure-Griffiths}, N.~M., {Dickey},
  J.~M., \& {Green}, A.~J. 2006, \apjs, 167, 230

\bibitem[{Henderson {et~al.}(1982)Henderson, Jackson, \& Kerr}]{henderson82}
Henderson, A.~P., Jackson, P.~D., \& Kerr, F.~J. 1982, \apj, 263, 116

\bibitem[{{Humphreys} \& {Kerr}(1974)}]{humphreys74}
{Humphreys}, R.~M. \& {Kerr}, F.~J. 1974, \apj, 194, 301

\bibitem[{{Kerr}(1962)}]{kerr62}
{Kerr}, F.~J. 1962, \mnras, 123, 327

\bibitem[{{Koo} \& {Kang}(2004)}]{koo04}
{Koo}, B.-C. \& {Kang}, J.-h. 2004, \mnras, 349, 983

\bibitem[{{Kulkarni} \& {Fich}(1985)}]{kulkarni85}
{Kulkarni}, S.~R. \& {Fich}, M. 1985, \apj, 289, 792

\bibitem[{{Kwee} {et~al.}(1954){Kwee}, {Muller}, \& {Westerhout}}]{kwee54}
{Kwee}, K.~K., {Muller}, C.~A., \& {Westerhout}, G. 1954, \bain, 12, 211

\bibitem[{{Lin} {et~al.}(1978){Lin}, {Yuan}, \& {Roberts}}]{lin78}
{Lin}, C.~C., {Yuan}, C., \& {Roberts}, Jr., W.~W. 1978, \aap, 69, 181

\bibitem[{{Lin} {et~al.}(1969){Lin}, {Yuan}, \& {Shu}}]{lin69}
{Lin}, C.~C., {Yuan}, C., \& {Shu}, F.~H. 1969, \apj, 155, 721

\bibitem[{{Liszt}(1983)}]{liszt83}
{Liszt}, H.~S. 1983, \apj, 275, 163

\bibitem[{{Luna} {et~al.}(2006){Luna}, {Bronfman}, {Carrasco}, \&
  {May}}]{luna06}
{Luna}, A., {Bronfman}, L., {Carrasco}, L., \& {May}, J. 2006, \apj, 641, 938

\bibitem[{Malhotra(1995)}]{malhotra95}
Malhotra, S. 1995, \apj, 448, 138

\bibitem[{{McClure-Griffiths} {et~al.}(2005){McClure-Griffiths}, {Dickey},
  {Gaensler}, {Green}, Haverkorn, \& Strasser}]{mcgriff05}
{McClure-Griffiths}, N.~M., {Dickey}, J.~M., {Gaensler}, B.~M., {Green}, A.~J.,
  Haverkorn, M., \& Strasser, S. 2005, \apjs, 158, 178

\bibitem[{Mignard(2000)}]{mignard00}
Mignard, F. 2000, \aap, 354, 522

\bibitem[{Press {et~al.}(1992)Press, Flannery, Teukolsky, \&
  Vetterling}]{press92}
Press, W.~H., Flannery, B.~P., Teukolsky, S.~A., \& Vetterling, W.~T. 1992,
  Numerical Recipes in Fortran ({Cambridge University Press})

\bibitem[{{Radhakrishnan} \& {Srinivasan}(1980)}]{rad80}
{Radhakrishnan}, V. \& {Srinivasan}, G. 1980, Journal of Astrophysics and
  Astronomy, 1, 47

\bibitem[{{Reid}(1993)}]{reid93}
{Reid}, M.~J. 1993, \araa, 31, 345

\bibitem[{{Roberts}(1969)}]{roberts69}
{Roberts}, W.~W. 1969, \apj, 158, 123

\bibitem[{{Rots}(1975)}]{rots75}
{Rots}, A.~H. 1975, \aap, 45, 43

\bibitem[{{Schmidt}(1956)}]{schmidt56}
{Schmidt}, M. 1956, \bain, 13, 15

\bibitem[{{Shane} \& {Bieger-Smith}(1966)}]{shane66}
{Shane}, W.~W. \& {Bieger-Smith}, G.~P. 1966, \bain, 18, 263

\bibitem[{{Sinha}(1978)}]{sinha78}
{Sinha}, R.~P. 1978, \aap, 69, 227

\bibitem[{{Stil} {et~al.}(2006){Stil}, {Taylor}, {Dickey}, {Kavars}, {Martin},
  {Rothwell}, {Boothroyd}, {Lockman}, \& {McClure-Griffiths}}]{stil06}
{Stil}, J.~M., {Taylor}, A.~R., {Dickey}, J.~M., {Kavars}, D.~W., {Martin},
  P.~G., {Rothwell}, T.~A., {Boothroyd}, A.~I., {Lockman}, F.~J., \&
  {McClure-Griffiths}, N.~M. 2006, \aj, 132, 1158

\bibitem[{{van Woerden} {et~al.}(1957){van Woerden}, {Rougoor}, \&
  {Oort}}]{vanwoerden57}
{van Woerden}, H., {Rougoor}, G.~W., \& {Oort}, J.~H. 1957, Comp.\ Rendu\, 244,
  1961

\bibitem[{{Visser}(1980)}]{visser80}
{Visser}, H.~C.~D. 1980, \aap, 88, 149

\bibitem[{{Weaver} \& {Williams}(1973)}]{weaver73}
{Weaver}, H. \& {Williams}, D.~R.~W. 1973, \aaps, 8, 1

\bibitem[{{Yuan}(1969)}]{yuan69}
{Yuan}, C. 1969, \apj, 158, 871

\end{thebibliography}

\normalsize

%---------------------------------------------
% Figures
%---------------------------------------------

%--------------------------------------------
% Tables
%--------------------------------------------
\clearpage

\begin{deluxetable}{lcc}
  \tabletypesize{\scriptsize} \tablecaption{Measured \HI\ terminal
    velocities, $v_{LSR}$, as a function of Galactic longitude, $l$,
    from the SGPS.  The third column gives the terminal velocities, $
    v_{mod}$, predicted by the rotation curve defined in Equation
    \ref{eq:rotfit}.  Note-- the full version of this table appears in the
    electronic edition of the Journal, an abbreviated form is
    presented here.
\label{tab:termvels}}
\tablewidth{0pt}
\tablehead{
  \colhead{$l$ (deg)} & \colhead{$v_{LSR}$ (\kms)} & \colhead{$v_{mod}$ (\kms)} 
}
\startdata
$339.695$  &   $-135.22$  &  $-133.00$   \\
$339.628$  &   $-135.14$  &  $-132.80$   \\
$339.562$  &   $-135.44$  &  $-132.61$   \\
$339.495$  &   $-136.17$  &  $-132.41$   \\
$339.428$  &   $-135.59$  &  $-132.22$   \\
$339.362$  &   $-137.87$  &  $-132.02$   \\
$339.295$  &   $-136.53$  &  $-131.83$   \\
$339.228$  &   $-135.63$  &  $-131.63$   \\
$339.162$  &   $-135.47$  &  $-131.44$   \\
$339.095$  &   $-134.11$  &  $-131.24$   \\
\enddata                   
\end{deluxetable}          

\begin{deluxetable}{lccccccc}
\tabletypesize{\scriptsize}
\tablecaption{Fitted values for velocity widths of two and
  three component error function fits to the terminal velocity curve
  averaged over the longitude range $272\arcdeg\leq l \leq 340\arcdeg$.
\label{tab:errfit}}
\tablewidth{0pt}
\tablehead{
  \colhead{} & \colhead{$v_o$ (\kms)} & \colhead{$a_1$ (K)} &
  \colhead{$\Delta v_1$ (\kms)} & \colhead{$a_2$ (K)} &
  \colhead{$\Delta v_2$ (\kms)} & \colhead{$a_3$ (K)} &
  \colhead{$\Delta v_3$ (\kms)}
}
\startdata
2-component fit & $7.7\pm0.3$ & $45.4\pm3.6$ & $8.6\pm0.3$ &
$8.1\pm0.8$ & $21.2\pm1.3$ & & \\
3-component fit & $7.0\pm0.1$ & $28.6\pm1.1$ & $6.3\pm0.1$ & $19.5\pm0.4$ &
$12.3\pm0.3$ & $3.1\pm0.1$ & $25.9\pm0.5$ \\
\enddata
\end{deluxetable}
\end{document}